\documentclass[twocolumn]{aastex62}
\pdfoutput=1 %for arXiv submission
\usepackage{amsmath,amstext}
\usepackage[T1]{fontenc}
\usepackage[figure,figure*]{hypcap}
\usepackage{natbib}
\usepackage[disable,colorinlistoftodos]{todonotes}

\usepackage{soul, CJK}

\graphicspath{{./}{fig/}}

\shorttitle{Tidal disruption flares are coming}
\shortauthors{van Velzen et al.}
\begin{document}

%Multi-wavelength observations of the first tidal disruption flare in ZTF (AT2018zr) and promising prospects for photometric selection}

\title{\Large \bf The first tidal disruption flare in ZTF: \\  from photometric selection to multi-wavelength characterization}

\begin{CJK*}{UTF8}{bkai}

\correspondingauthor{Sjoert van Velzen}
\email{sjoert@umd.edu}

\author[0000-0002-3859-8074]{Sjoert van Velzen}
\affiliation{Department of Astronomy, University of Maryland, College Park, MD 20742}
\affiliation{Center for Cosmology and Particle Physics, New York University, NY 10003}

\author[0000-0003-3703-5154]{Suvi Gezari}
\affiliation{Department of Astronomy, University of Maryland, College Park, MD  20742, USA}
\affiliation{Joint Space-Science Institute, University of Maryland, College Park, MD 20742, USA}

\author[0000-0003-1673-970X]{S. Bradley Cenko}
\affiliation{Astrophysics Science Division, NASA Goddard Space Flight Center, MC 661, Greenbelt, MD 20771, USA}
\affiliation{Joint Space-Science Institute, University of Maryland, College Park, MD 20742, USA}

\author[0000-0003-0172-0854]{Erin Kara}
\affiliation{Department of Astronomy, University of Maryland, College Park, MD  20742, USA}
\affiliation{X-ray Astrophysics Laboratory, NASA/Goddard Space Flight Center, Greenbelt, MD 20771, USA.}
\affiliation{Joint Space-Science Institute, University of Maryland, College Park, MD 20742, USA}

\author[0000-0003-3124-2814]{James C. A. Miller-Jones}
\affiliation{ICRAR -- Curtin University, GPO Box U1987, Perth, WA 6845, Australia}

\author{Tiara Hung}
\affiliation{Department of Astronomy, University of Maryland, College Park, MD 20742, USA}

\author{Joe Bright}
\affiliation{Department of Physics, University of Oxford, Denys Wilkinson Building, Keble Road, Oxford OX1 3RH, UK}

\author[0000-0002-6485-2259]{Nathaniel Roth}
\affiliation{Department of Astronomy, University of Maryland, College Park, MD  20742, USA}
\affiliation{Joint Space-Science Institute, University of Maryland, College Park, MD 20742, USA}

\author[0000-0003-0901-1606]{Nadejda Blagorodnova}
\affiliation{Division of Physics, Mathematics, and Astronomy, California Institute of Technology, Pasadena, CA 91125, USA}

\author{Daniela Huppenkothen}
\affiliation{DIRAC Institute, Department of Astronomy, University of Washington, 3910 15th Avenue NE, Seattle, WA 98195, USA}

\author[0000-0003-1710-9339]{Lin Yan}
\affiliation{The Caltech Optical Observatories, California Institute of Technology, Pasadena, CA 91125, USA}

\author{Eran Ofek}
\affiliation{Benoziyo Center for Astrophysics and the Helen Kimmel Center for Planetary Science, Weizmann Institute of Science, 76100 Rehovot, Israel}

\author{Jesper Sollerman}
\affiliation{The Oskar Klein Centre \& Department of Astronomy, Stockholm University, AlbaNova, SE-106 91 Stockholm, Sweden}

\author[0000-0001-9676-730X]{Sara~Frederick}
\affiliation{Department of Astronomy, University of Maryland, College Park, MD 20742, USA}

\author{Charlotte Ward}
\affiliation{Department of Astronomy, University of Maryland, College Park, MD 20742, USA}

\author[0000-0002-3168-0139]{Matthew J. Graham}
\affiliation{Division of Physics, Mathematics, and Astronomy, California Institute of Technology, Pasadena, CA 91125, USA}

\author{Rob Fender}
\affiliation{Department of Physics, University of Oxford, Denys Wilkinson Building, Keble Road, Oxford OX1 3RH, UK}

\author{Mansi M. Kasliwal}
\affiliation{Division of Physics, Mathematics, and Astronomy, California Institute of Technology, Pasadena, CA 91125, USA}

\author{Chris Canella}
\affiliation{Division of Physics, Mathematics, and Astronomy, California Institute of Technology, Pasadena, CA 91125, USA}

\author{Robert Stein}
\affiliation{Deutsches Elektronensynchrotron, Platanenallee 6, D-15738, Zeuthen, Germany}

\author{Matteo Giomi}
\affiliation{Institute of Physics, Humboldt-Universit\"at zu Berlin, Newtonstr. 15, D-12489 Berlin, Germany}

\author{Valery Brinnel}
\affiliation{Institute of Physics, Humboldt-Universit\"at zu Berlin, Newtonstr. 15, 12489 Berlin, Germany}

\author{Jakob van Santen}
\affiliation{Deutsches Elektronensynchrotron, Platanenallee 6, D-15738, Zeuthen, Germany}

\author{Jakob Nordin}
\affiliation{Institute of Physics, Humboldt-Universit\"at zu Berlin, Newtonstr. 15, 12489 Berlin, Germany}

%\nocollaboration{}

%---
% ZTF opt-in authors below (alphabetical)
% ----

\author[0000-0001-8018-5348]{Eric C. Bellm}
\affiliation{DIRAC Institute, Department of Astronomy, University of Washington, 3910 15th Avenue NE, Seattle, WA 98195, USA}

\author{Richard Dekany}
\affiliation{Caltech Optical Observatories, California Institute of Technology, Pasadena, CA 91125, USA}

\author{Christoffer Fremling}
\affiliation{Division of Physics, Mathematics, and Astronomy, California Institute of Technology, Pasadena, CA 91125, USA}

\author{V. Zach Golkhou} 
\affiliation{DIRAC Institute, Department of Astronomy, University of Washington, 3910 15th Avenue NE, Seattle, WA 98195, USA} 
\affiliation{The eScience Institute, University of Washington, Seattle, WA 98195, USA}

\author[0000-0002-6540-1484]{Thomas Kupfer}
\affiliation{Kavli Institute for Theoretical Physics, University of California, Santa Barbara, CA 93106, USA}
\affiliation{Department of Physics, University of California, Santa Barbara, CA 93106, USA}
\affiliation{Division of Physics, Mathematics, and Astronomy, California Institute of Technology, Pasadena, CA 91125, USA}

\author[0000-0001-5390-8563]{Shrinivas R. Kulkarni}
\affiliation{Division of Physics, Mathematics, and Astronomy, California Institute of Technology, Pasadena, CA 91125, USA}

\author{Russ R. Laher}
\affil{Infrared Processing and Analysis Center, California Institute of Technology, Pasadena, CA 91125, USA.}

\author[0000-0003-2242-0244]{Ashish Mahabal}
\affiliation{Division of Physics, Mathematics, and Astronomy, California Institute of Technology, Pasadena, CA 91125, USA}
\affiliation{Center for Data Driven Discovery, California Institute of Technology, Pasadena, CA 91125, USA}

\author[0000-0002-8532-9395]{Frank J. Masci}
\affiliation{Infrared Processing and Analysis Center, California Institute of Technology, MS 100-22, Pasadena, CA 91125, USA}

\author[0000-0001-9515-478X]{Adam A. Miller}
\affil{Center for Interdisciplinary Exploration and Research in Astrophysics (CIERA) and Department of Physics and Astronomy, Northwestern University, 2145 Sheridan Road, Evanston, IL 60208, USA}
\affil{The Adler Planetarium, Chicago, IL 60605, USA}

\author{James D. Neill}
\affiliation{Division of Physics, Mathematics, and Astronomy, California Institute of Technology, Pasadena, CA 91125, USA}

\author{Reed Riddle}
\affiliation{Caltech Optical Observatories, California Institute of Technology, Pasadena, CA 91125, USA}

\author{Mickael Rigault}
\affiliation{Université Clermont Auvergne, CNRS/IN2P3, 
Laboratoire de Physique de Clermont, F-63000 Clermont-Ferrand, France.} 

\author[0000-0001-7648-4142]{Ben Rusholme}
\affiliation{Infrared Processing and Analysis Center, California Institute of Technology, MS 100-22, Pasadena, CA 91125, USA}

\author[0000-0001-6753-1488]{Maayane T. Soumagnac}
\affiliation{Department of Particle Physics and Astrophysics, Weizmann Institute of Science 
234 Herzl St., Rehovot, 76100, Israel}

\author[0000-0001-6584-6945]{Yutaro Tachibana (優太朗橘)}
\affil{Department of Physics, Tokyo Institute of Technology, 2-12-1 Ookayama, Meguro-ku, Tokyo 152-8551,
Japan}
\affil{Department of Physics, Math, and Astronomy, California Institute of Technology, Pasadena, CA 91125, USA}

%\collaboration{(ZTF collaboration)} % we used an opt-in process, so the author above are not the entire ZTF collaboration

\begin{abstract}
We present Zwicky Transient Facility (ZTF) observations of the tidal disruption flare AT2018zr/PS18kh reported by Holoien et al. and detected during ZTF commissioning. The ZTF light curve of the tidal disruption event (TDE) samples the rise-to-peak exceptionally well, with 50~days of $g$- and $r$-band detections before the time of maximum light. We also present our multi-wavelength follow-up observations, including the detection of a thermal ($kT\approx 100$~eV) X-ray source that is two orders of magnitude fainter than the contemporaneous optical/UV blackbody luminosity, and a stringent upper limit to the radio emission. 
We use observations of 128 known active galactic nuclei (AGN) to assess the quality of the ZTF astrometry, finding a median host-flare distance of 0$\farcs$2 for genuine nuclear flares. Using ZTF observations of variability from known AGN and supernovae we show how these sources can be separated from TDEs. A combination of light-curve shape, color, and location in the host galaxy can be used to select a clean TDE sample from multi-band optical surveys such as ZTF or LSST. 
\end{abstract}


\section{Introduction} \label{sec:intro}
Stars that pass within the tidal radius of a supermassive black hole are disrupted and a sizable fraction of the resulting stellar debris gets accreted onto the black hole. When this disruption occurs outside the black hole event horizon \citep{Hills75}, the result is a luminous flare of thermal emission \citep{Rees88}. These stellar tidal disruption flares provide a unique tool to study black hole accretion and jet formation \citep[e.g.,][]{Giannios11,vanVelzen11,Tchekhovskoy13,Coughlin14,Piran15,PashamvanVelzen17}.

Optical transient surveys currently dominate the discovery of tidal disruption events (TDEs); about a dozen candidates have been found to date \citep[for a recent compilation see][]{vanVelzen18}. All-sky X-ray surveys provide a second avenue for discovery \citep[e.g.,][]{Saxton17}. By late 2019, the eROSITA mission \citep{Merloni12} should significantly increase the number of TDEs discovered via their X-ray emission \citep{Khabibullin14a}.

The observed blackbody radii of known TDEs \citep{Gezari09} suggest the soft X-ray photons of these flares originate from the inner part of a newly formed accretion disk ($\sim 10^{11}$~cm), while the optical photons are produced at much larger radii, $\sim 10^{14}$~cm. Only a handful of optically selected TDEs have received sensitive X-ray follow-up observations within the first few months of discovery. So far, every case has been different; optically selected TDEs can be X-ray faint \citep{Gezari12}, or have $L_{\rm opt}/L_{X}\sim 1$ \citep{Holoien16}, or even show a decreasing optical-to-X-ray ratio \citep{Gezari17}.

Unification of the optical and X-ray properties of TDEs is possible if the optical emission is powered by shocks from intersecting stellar debris streams \citep{Piran15} and the X-ray photons are produced when parts of the stream get deflected toward a few gravitational radii and accreted \citep{Shiokawa15,Krolik16}. In this scenario, TDEs with low X-ray luminosities can be explained as inefficiencies in this circularization process. If instead most of the stellar debris is able to rapidly form an accretion disk \citep[][]{Bonnerot16,
Hayasaki16}, the X-rays from the disk have to be reprocessed at larger radii \citep[e.g.,][]{LoebUlmer97,Bogdanovic04,strubbe_quataert09,Guillochon14} to yield the observed optical emission. When the reprocessing layer is optically thick to X-rays, TDE unification is established via orientation \citep[e.g.,][]{Metzger16,Auchettl16,Dai18}. 

%Discriminating between the different unification proposals might be possible by comparing the X-ray and optical TDE rate (e.g., if the reprocessing picture is correct, the ratio of the X-ray/optical rate yields the covering factor of the region that absorbs the X-ray photons). However this requires a larger body of TDEs since the current rate estimates are dominated by Poisson uncertainties \citep{vanVelzen14,Holoien16,Hung18}. Moreover, rates from different X-ray surveys \citep{Donley02,Esquej07} are discrepant by a factor $\approx 70$ ---although this discrepancy can be fixed by a more uniform analysis of the X-ray TDEs (Saxton et al. 2018 in prep). 

Insight into the emission mechanism of TDEs can be gained from detailed observations of individual events. A lag of the X-ray emission in a cross-correlation analysis \citep[][]{Pasham17} and a decrease of $L_{\rm opt}/L_X$ with time \citep{Gezari17a} have both been interpreted as evidence against a reprocessing scenario. However,  these conclusions are not definitive since the X-ray diversity of TDEs has not yet been mapped out. 

% this first TDE candidate is encouraging evidence that this survey will reach the anticipated performance. 
% which brings us to the topic of this paper: a new TDE detected in commissioning data from the Zwicky Transient Facility (ZTF)

Clearly more TDEs with multi-wavelength observations are needed to make progress. As discussed in \citet{Hung18}, ZTF has the potential to significantly increase the TDE detection rate. Like the Palomar Transient Factory \citep[PTF;][]{Law09,Rau09}, ZTF uses the Samuel Oschin 48" Schmidt telescope at Palomar Observatory. The biggest improvement over PTF is the 47~deg$^2$ field-of-view of the ZTF camera \citep{Bellm19}. The public Northern Sky Survey of ZTF  \citep{Graham19} began on 2018 March 17, and covers the entire visible sky from Palomar in both the $g$ and $r$ bands every three nights (the Galactic Plane, $|b| < 7^\circ$, is covered with a one night cadence) to a typical depth of 20.5 mag. 
Using ZTF images, a stream of alerts \citep{Patterson19} containing transients and variable sources is generated by IPAC \citep{Masci19}. Besides the essential photometric information, this stream contains value-added products such as the quality of the subtraction \citep{Mahabal19} and the probability that the alert is associated with a star versus a galaxy \citep{Tachibana18}. 
%Commissioning of ZTF began in late 2017, the first scientific results followed in early 2018 \citep{Kulkarni18}. 

This paper is organized as follows. In Section~\ref{sec:results} we present our observations of \mbox{AT2018zr}/PS18kh \citep{Holoien18a}, the first TDE with ZTF observations. This source was discovered in Pan-STARRS \citep{Chambers16} imaging data; both ATLAS \citep{Tonry18} and ASAS-SN \citep{Shappee14} also obtained detections, see \citet{Holoien18a}. Here we present several new observations of this latest TDE: the ZTF light curve,  {\em XMM-Newton} X-ray spectra, and VLA/AMI radio observations. The results from our HST campaign of UV spectroscopy and ground-based optical spectroscopic monitoring will be presented separately (T. Hung et al., in prep.). In Section~\ref{sec:compare} we compare \mbox{AT2018zr} to previous TDEs, supernovae (SNe), and AGN. In Section~\ref{sec:offset} we show the astrometric quality of ZTF data for nuclear transients. In Section~\ref{sec:dics} we discuss the results.

\begin{figure}
\centering
\includegraphics[width=0.45 \textwidth, trim=4mm 3mm 4mm 6mm, clip]{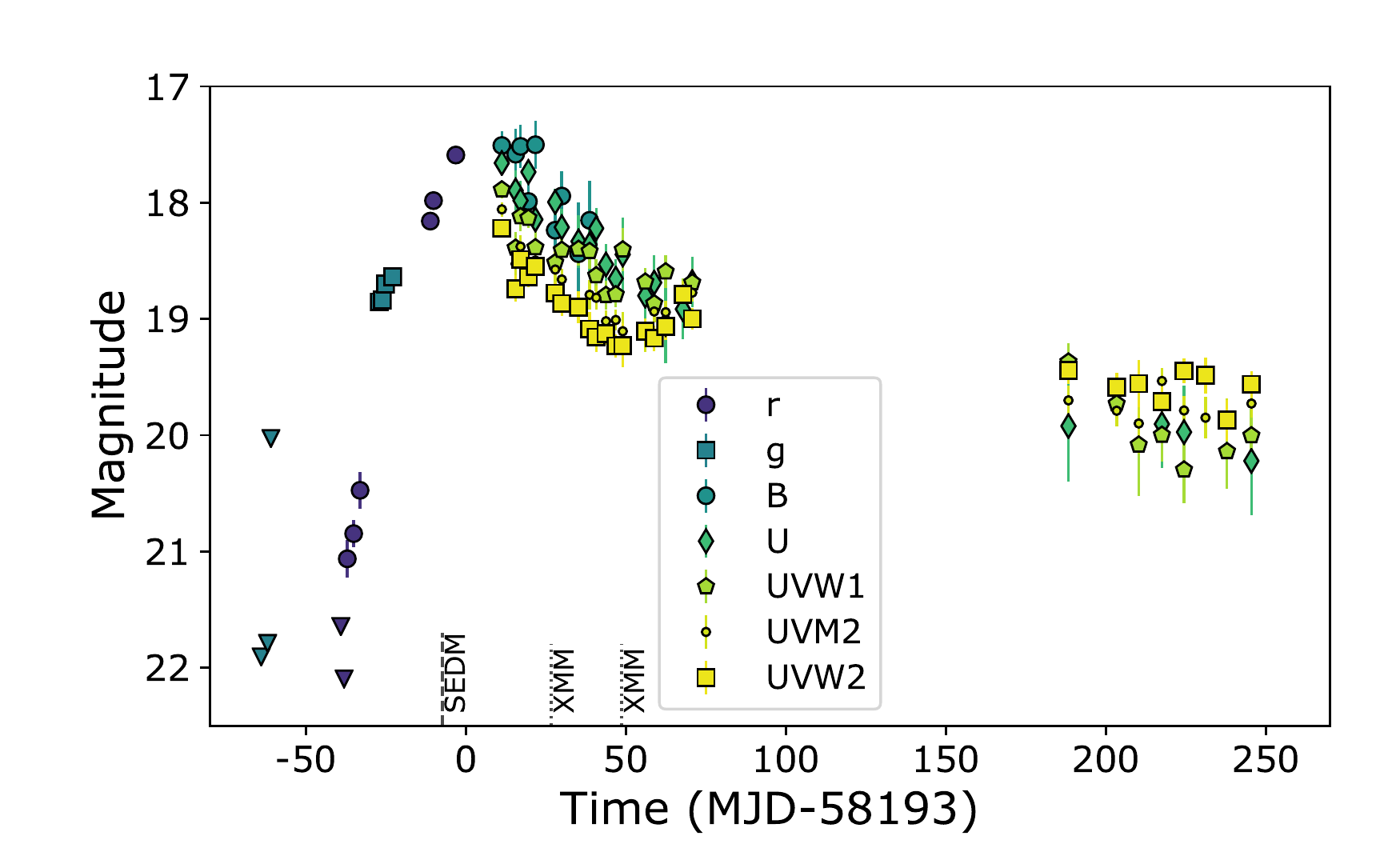}
\caption{ZTF and {\it Swift}/UVOT light curve. The dashed line indicates the time of the first SEDM spectrum, while the dotted lines label the times of XMM X-ray observations. Triangles denote 5$\sigma$ upper limits to the flux. }\label{fig:lc}
\end{figure}

We adopt a flat cosmology with $\Omega_\Lambda=0.7$ and $H_0=70~{\rm km}\,{\rm s}^{-1}{\rm Mpc}^{-1}$. All magnitudes are reported in the AB system \citep{oke74}. %For the redshift of NedStark/PS18kh we use $z=0.074$ \citep{Holoien18a}.

\begin{figure*}
\centering
\includegraphics[width=0.85 \textwidth, trim=3mm 1mm 4mm 6mm, clip]{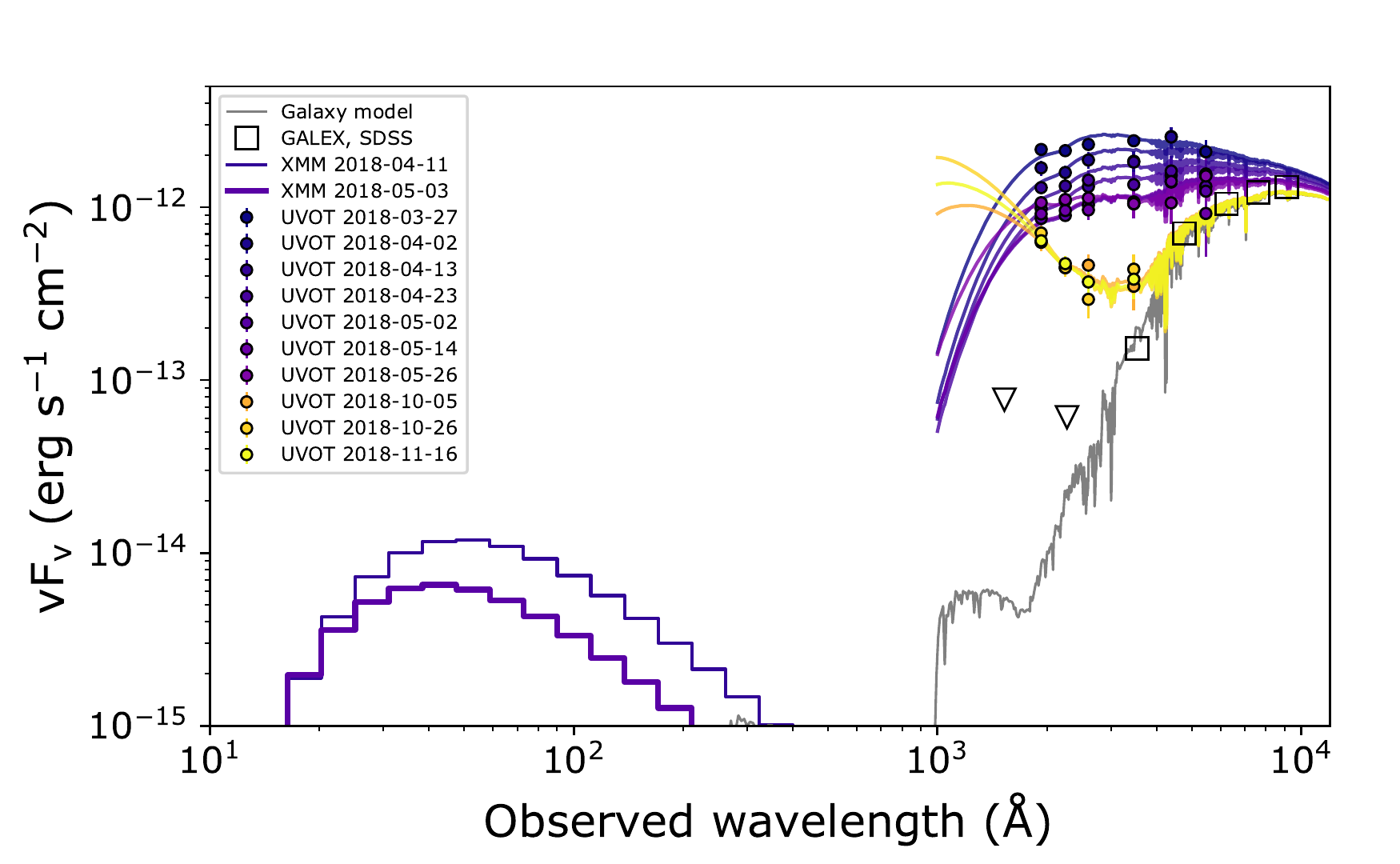}
\caption{Spectral energy distribution. Gray open symbols indicate the host galaxy photometry (from SDSS and GALEX, obtained before the flare) and the best-fit synthetic galaxy spectrum is shown by the thin gray line. The circles show a subset of the UVOT monitoring observations and the corresponding best-fit blackbody spectrum. The unfolded X-ray spectra obtained from the two epochs of {\em XMM-Newton} observations are also shown.}\label{fig:sed}
\end{figure*}

%----
\section{Observations}\label{sec:results}
%---

\subsection{Selection of nuclear flares in ZTF  data}\label{sec:nuc_select}
During the commissioning phase of the ZTF camera and IPAC alerts pipeline we assessed the quality of the alert stream, focusing on nuclear transients. A transient was considered nuclear when it had at least one detection with a distance between the location of the source in the reference frame and the location of the transient that was smaller than 0\farcs6. We also required a match within 1" of a known Pan-STARRS \citep{Chambers16} galaxy, selected using a star-galaxy score \citep{Tachibana18} of \verb sgscore $<0.3$. To remove sources with a very small flux increase, we also required that the point spread function (PSF) magnitude in the difference image (\verb magpsf , see \citealt{Masci19}) and the PSF magnitude of the source in the reference image (\verb magnr ) obey the relation: \verb magpsf $-$ \verb magnr $<$ 1.5 mag. 
To apply these filters and to obtain visual confirmation of the alerts we used the GROWTH Marshal \citep{Kasliwal19}.
%Cuts on other parameters (e.g., on the real-bogus parameter, ) were not kept constant since the pipeline that computes these parameters was still in flux during the commission phase. 

The objective of our commissioning effort was to understand the quality of the astrometry of nuclear transients; these results are presented in Section~\ref{sec:offset}.  We also obtained spectroscopic follow-up for a subset of nuclear transients, which led to the identification of the transient AT2018zr as a TDE candidate.   

\subsection{Brief history of AT2018zr}
On 2018 March 3, the Pan-STARRS survey discovered transient PS18kh with an $i$-band magnitude of $18.63$, and registered it on the Transient Name Server (TNS\footnote{\url{https://wis-tns.weizmann.ac.il}}) as AT2018zr the following day. On March 24, \citet[][ATel 11473]{Tucker18} reported a potential TDE classification for this source based on multiple spectroscopic observations that showed a very blue continuum and some evidence for broad Balmer emission lines. Photometry from Pan-STARRS, ASASSN, and ATLAS, as well as further spectroscopic observations identifying AT2018zr as a TDE, are presented in \citet{Holoien18a}.

On 2018 March 6, the source ZTF18aabtxvd\footnote{We internally nicknamed this source ZTF-NedStark.} was identified as a nuclear transient by our ZTF alert pipeline. We obtained a spectroscopic follow-up observation using SEDM \citep{Blagorodnova18} on 2018, March 7, which showed a nearly featureless blue continuum. We triggered HST UV spectroscopic observations on 2018, March 27. We report these observations, plus our spectroscopic monitoring with several ground-based telescopes in Hung et al. 2018, in prep, from which we determine a redshift of $z=0.071$.  

Upon further investigation we noticed the ZTF reference frame was contaminated with light from the transient, which prohibited an earlier detection; after rebuilding the reference images and applying an image subtraction algorithm \citep{Zackay16} that is similar to the one used in the IPAC pipeline, we found the first ZTF detection was on 2018 February 7 (Fig.~\ref{fig:lc}). 

\subsection{Optical/UV observations of AT2018zr}\label{sec:opticalUV}
Observations with the Neil Gehrels Swift Observatory \citep[{\it Swift};][]{Gehrels04} started on 2018 March 27 (PI: Holoien). We extracted the UVOT \citep{Roming05,Poole08} flux with the help of the \verb uvotsource  task, using an aperture radius of 5~arcsec. 

The flux of the host galaxy in the UVOT bands was estimated by fitting a synthetic galaxy spectrum \citep{Conroy09,Conroy10} to the SDSS model magnitudes \citep{stoughton02,Ahn14}, see Table~\ref{tab:host_syn}. To construct the synthetic galaxy spectrum we adopt the default assumptions for the stellar parameters: a \citealt{Kroupa01} initial mass function with stellar masses $0.08<M_{\rm star}/M_\odot<150$; Padova isochrones and MILES spectral library \citep{Vazdekis10}. We assume an exponentially declining star formation rate ($\exp(-t_{\rm age}/\tau_{\rm SF})$, with $t_{\rm age}$ and $\tau_{\rm SF}$ as free parameters). We account for Galactic extinction by applying the \citet{cardelli89} extinction law with $R_V=3.1$ to the model spectrum. We also allow for extinction in the TDE host galaxy by modifying the model spectrum using a \citet{Calzetti00} extinction law. The best-fit parameters for the  formation history are $t_{\rm age}=9.8$~Gyr and $\tau_{\rm SF}=1$~Gyr; the total stellar mass of the galaxy inferred for this model is $5\times 10^9\,M_\odot$.  

Besides the {\it Swift}/UVOT and ZTF photometry, our light curve also includes P60/SEDM photometric data, host-subtracted using SDSS reference images \citep{Fremling16}.

We correct the difference magnitude in each band for Galactic extinction, $E(B-V)=0.040$~mag  \citep{Schlegel98}, again assuming a \cite{cardelli89} extinction law with $R_V=3.1$. The resulting light curve is shown in Fig.~\ref{fig:lc} and the photometry is available in Table~\ref{tab:photo}.

By adding a blackbody spectrum to the synthetic host galaxy spectrum we find the best-fit temperature of the flare (Fig.~\ref{fig:sed}). During the first 40 days of {\it Swift} monitoring the mean temperature was $1.4\times 10^4$~K, and appears constant with an rms of only $0.1\times 10^4$~K. Starting near May 11, 2018, until the last observation before the source moved out of the {\it Swift} visibility window (May 29, 2018), the temperature increased by a factor of $\approx 1.5$ \citep[see also][]{Holoien18a}. The most recent observations, obtained when the source became visible again to {\it Swift}, show a large increase of the blackbody temperature; the UVOT observations are now on the Rayleigh Jeans tail of the SED and we obtain a lower limit to the temperature, $T\gtrsim 5 \times 10^4$~K.
%The temperature increase is also visible in the light curve (Fig.~\ref{fig:lc}) as an increase of the flux in the bluest UVOT filter (UVW2). 

We searched for time lags between the optical and UV measurements taken with {\it Swift} using cross-correlations, following the procedure in \citet{Peterson98}. We first correct the light curves with a simple linear trend using a maximum-likelihood approach, and then use linear interpolation between data points in order to sample both UV and optical light curves on the same grid. We find no significant time lags between any combination of the the UV and optical bands at the 95\% confidence level. % the number of data points is small, so the statistical power in the tests we used is weak. 

To search for outbursts in the years prior to the ZTF detection of AT2018zr, we applied a forced photometry method to the difference images from PTF and iPTF \citep{Masci17}. We obtained 61 images, clustered at 6, 4, and 1 year before the current peak of the light curve. No prior variability was detected to a typical $R$-band magnitude $m>20.6$  (5$\sigma$). 

Using our ZTF observations of AT2018zr, we measure a mean angular distance between the host galaxy center and the flare of  0\farcs12, or 162~pc. The rms of the offset, combining 46 offset measurements in Right Ascension and Declination and both $r$- and $g$-band detections, is 0\farcs25. We thus conclude that the position of the flare is consistent with originating from the center of its host galaxy. Indeed, as we will see in Section~\ref{sec:offset}, flares from AGN---which are expected to originate from the center of their galaxy---have a similar mean host-flare distance.

\subsection{X-ray observations of AT2018zr}
X-ray follow-up observations of AT2018zr were obtained using {\em XMM-Newton}  (program 082204, PI: Gezari).
Two epochs of {\em XMM-Newton} observations of the source were taken on 2018 April 11, and 2018 May 3 (see Table~\ref{tab:xmm} for details). We reduced the {\em XMM-Newton}/pn data using the {\em XMM-Newton} Science Analysis System (SAS) and the newest calibration files. We started with the observation data files (ODFs) and followed standard procedures. Events were filtered with the conditions \verb PATTERN<= 4  and \verb FLAG==0 . We checked for high background flares (of which there were none). The source and background extraction regions are circular regions of radius 35". The response matrices were produced using {\sc rmfgen} and {\sc arfgen} in SAS. Spectral fitting was performed using
{\sc XSPEC} v12.10 \citep{Arnaud96} with the c-statistic. 

\begin{figure}
\centering
\includegraphics[trim=0pt 3pt 0pt -5pt, width=0.95\columnwidth]{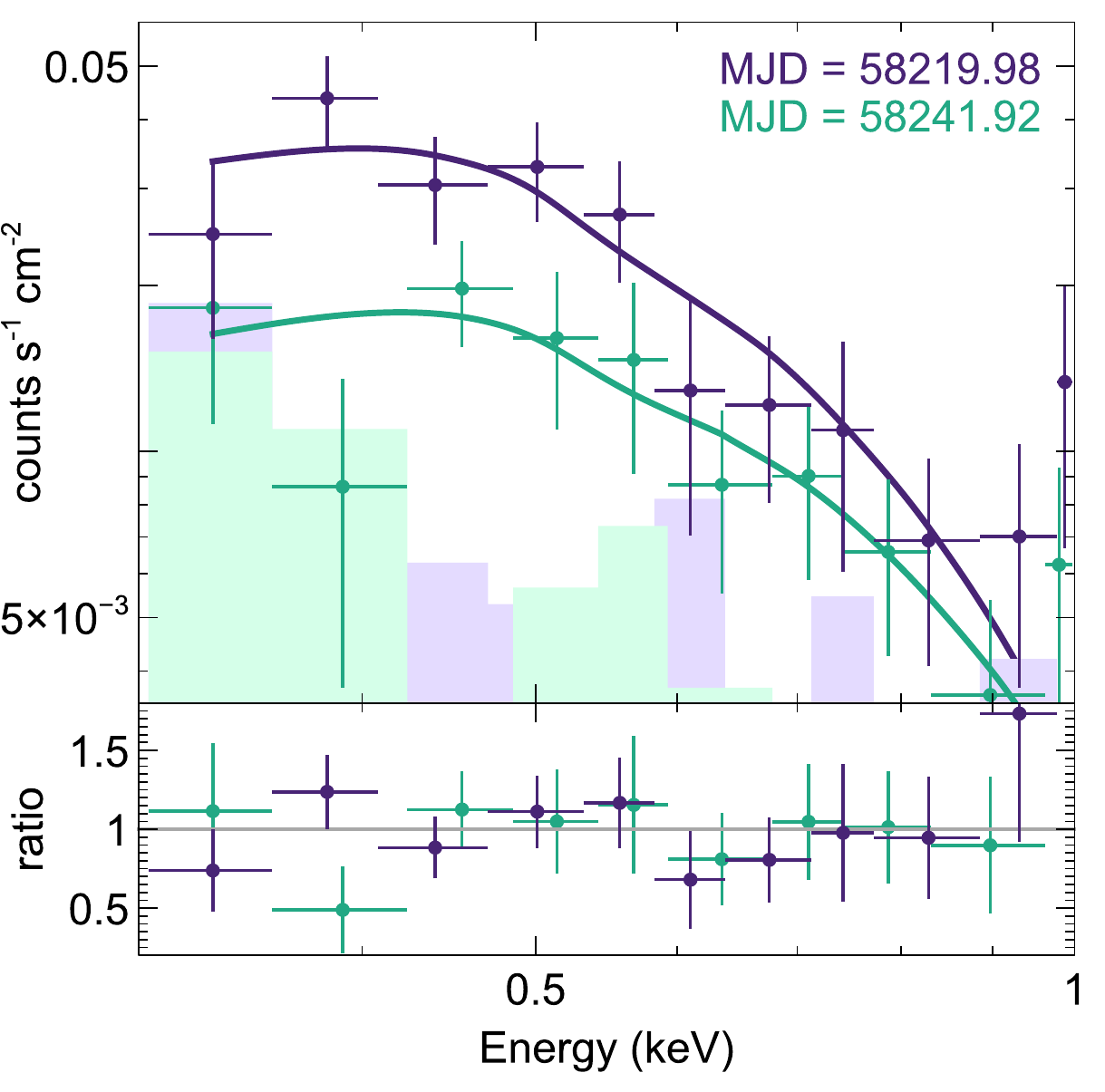}
\caption{{\em Top:} The {\em XMM-Newton} spectra for the two epochs. The source spectrum is shown in by the solid lines and the background is shown as the shaded regions. Both epochs are well described by a single blackbody component with Galactic absorption (solid lines). {\em Bottom:} The ratio of the spectra to the best fit model.}\label{fig:xmm_spec}
\end{figure}

Both spectra are well described by a single blackbody component ($T\approx 100$~eV) and Galactic absorption ($N_{\rm H} = 4.4 \times 10^{20}$ cm$^{-2}$; \citealt{Kalberla2005}). No additional absorption (at the redshift of the source) was required. The 0.3-1~keV luminosity of the first epoch is $L_X=2\times 10^{41} \, {\rm erg}\,{\rm s}^{-1}$. The temperature of the thermal component remained constant between the two observations, though the flux decreased by a factor of 2. In Table~\ref{tab:xmm} we list the full details of the spectral parameters.  

The results from our XMM analysis are in mild conflict with the non-thermal spectrum (photon index $\Gamma =3 \pm 1$) reported  by \citet{Holoien18a} using the {\it Swift}/XRT data. The XRT measurements overlap with our {\em XMM-Newton} epochs, but have a lower signal-to-noise ratio and less spectral coverage at low energies. Given these limitations, a thermal spectrum is easily mistaken for a steep power-law. Indeed, the XRT flux reported by \citet{Holoien18a} is consistent with the {\em XMM-Newton} flux. Given the superior quality of the XMM observations, we conclude that the thermal nature of the X-ray spectrum of AT2018zr is firmly established. 

At late-time, $175-250$~days after peak, no XMM observations are available. To estimate the X-ray flux we binned the {\it Swift}/XRT observations of these epochs, yielding 14 photons in 77~ks of time on-source. The signal in this detection is not sufficient to measure the spectrum---although we note that the majority of the counts originate from the low-energy (0.3--1~keV) channels, indicating the spectrum has remained soft. To estimate the flux from the binned XRT observations we use a blackbody spectrum with a fixed temperature of 100~keV. We find that the late-time X-ray flux has not decreased significantly; the flux derived from the binned XRT observations is consistent with flux from the last  {\em XMM-Newton} epoch; see Table~\ref{tab:xmm}.

\subsection{Radio upper limits of AT2018zr}
Radio observations of AT2018zr were obtained using the Arcminute Microkelvin Imager Large Array (AMI-LA; \citealt{zwartAMI2008,hickishAMI2017}) and the Karl~G. Jansky Very Large Array (VLA; program 18A-373, PI: van~Velzen). AMI observed on 2018 March 28, followed by the VLA on 2018 March 30 and April 28. The AMI data reduction was performed using the custom calibration pipeline \textsc{reduce\_dc} (e.g. \citealt{perrott2013}) with 3C\,286 used as the primary calibrator, and J0745+3142 as the secondary calibrator. The same calibrators were used for both VLA observations. For the VLA data analysis, we made use of the NRAO pipeline products and flagged a few additional spectral channels after manual inspection for radio frequency interference. The calibrated visibilities were imaged using the Common Astronomy Software Application \citep[CASA;][]{McMullin07} task \texttt{clean}, with natural weighting. 

The source was not detected in any of our radio observations.  The rms was determined from a source-free region adjacent to the target position. Our most sensitive observation was the first VLA epoch, which yields a 3$\sigma$ upper limit to the 10~GHz radio luminosity (defined as $L_{\rm r} = 4\pi d^2 \nu S_{\nu}$) of $<1\times 10^{37}$~erg~s$^{-1}$. Full details are listed in Table~\ref{tab:janksy}.

\begin{figure}
\centering
\includegraphics[width=0.45 \textwidth, trim=3mm 3mm 4mm 6mm, clip]{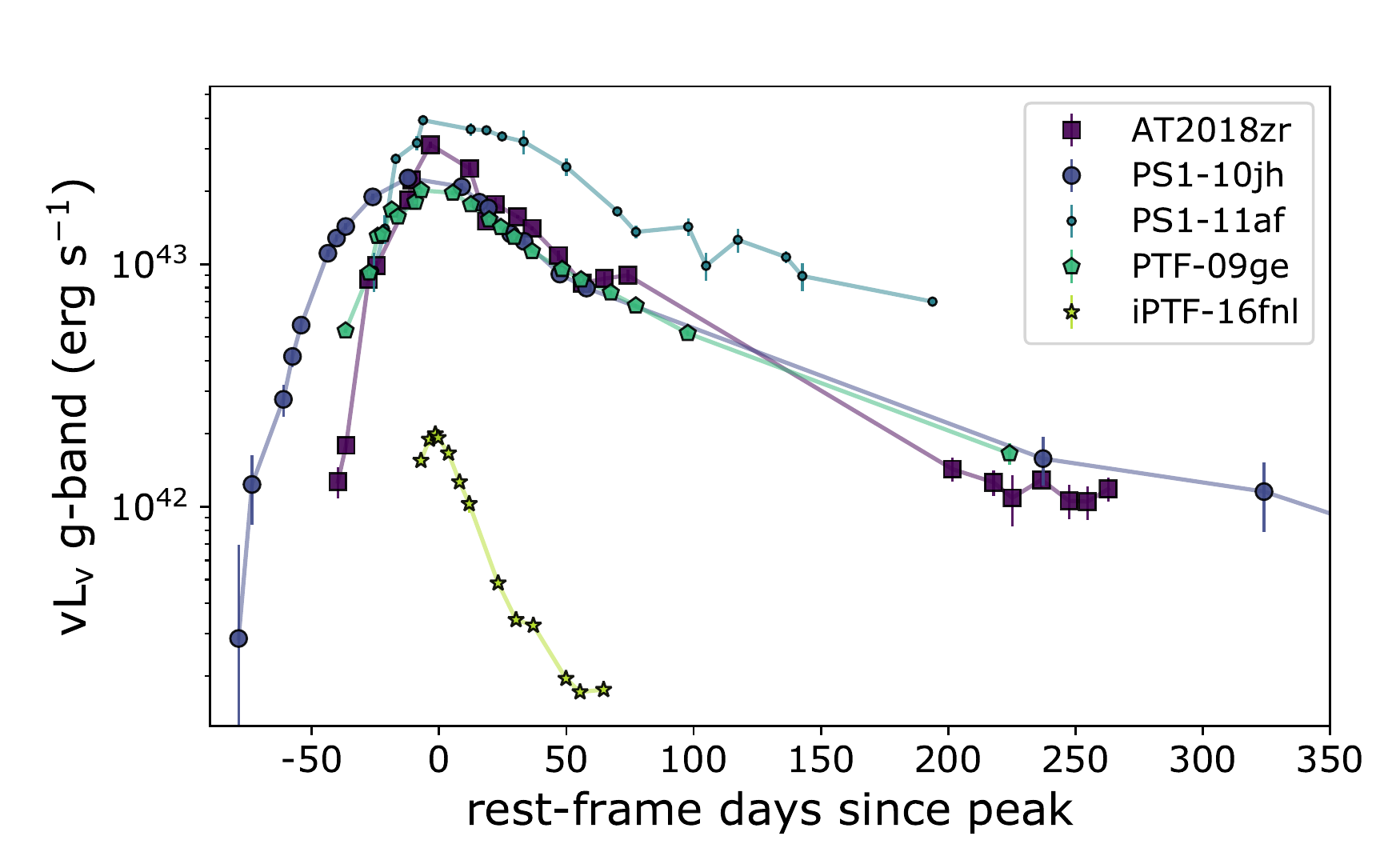}\\
\includegraphics[width=0.45 \textwidth, trim=3mm 3mm 4mm 6mm, clip]{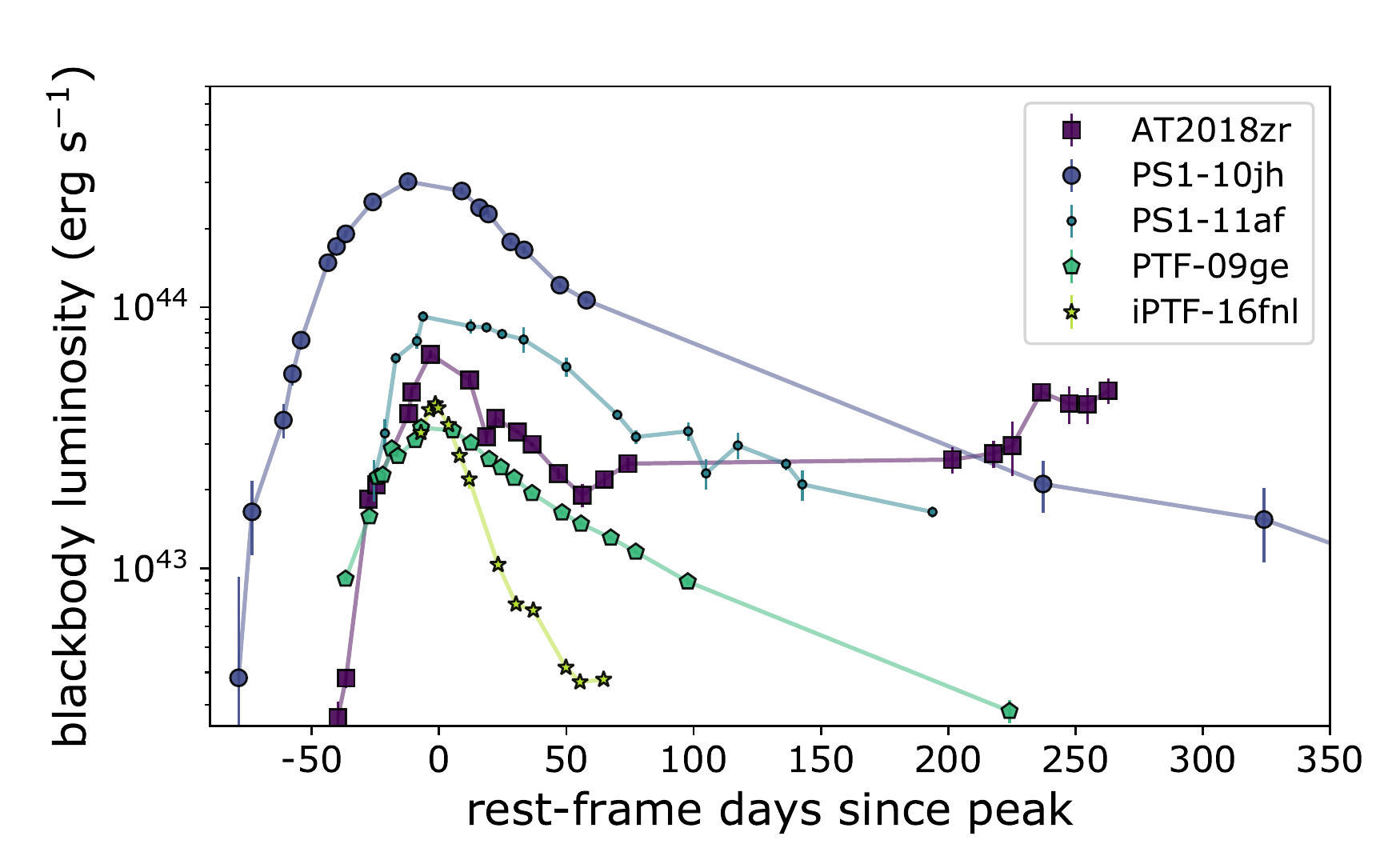}
\caption{Light curves of the five optical TDEs that have a resolved rise-to-peak. We show both the rest-frame \mbox{$g$-band} luminosity (top) and the blackbody luminosity (bottom). Note the difference of the vertical axis scale between the two figures due to the bolometric correction from the optical luminosity to the blackbody luminosity. }\label{fig:lcs}
\end{figure}

\begin{figure}
\centering
\includegraphics[width=0.45 \textwidth, trim=3mm 3mm 4mm 6mm, clip]{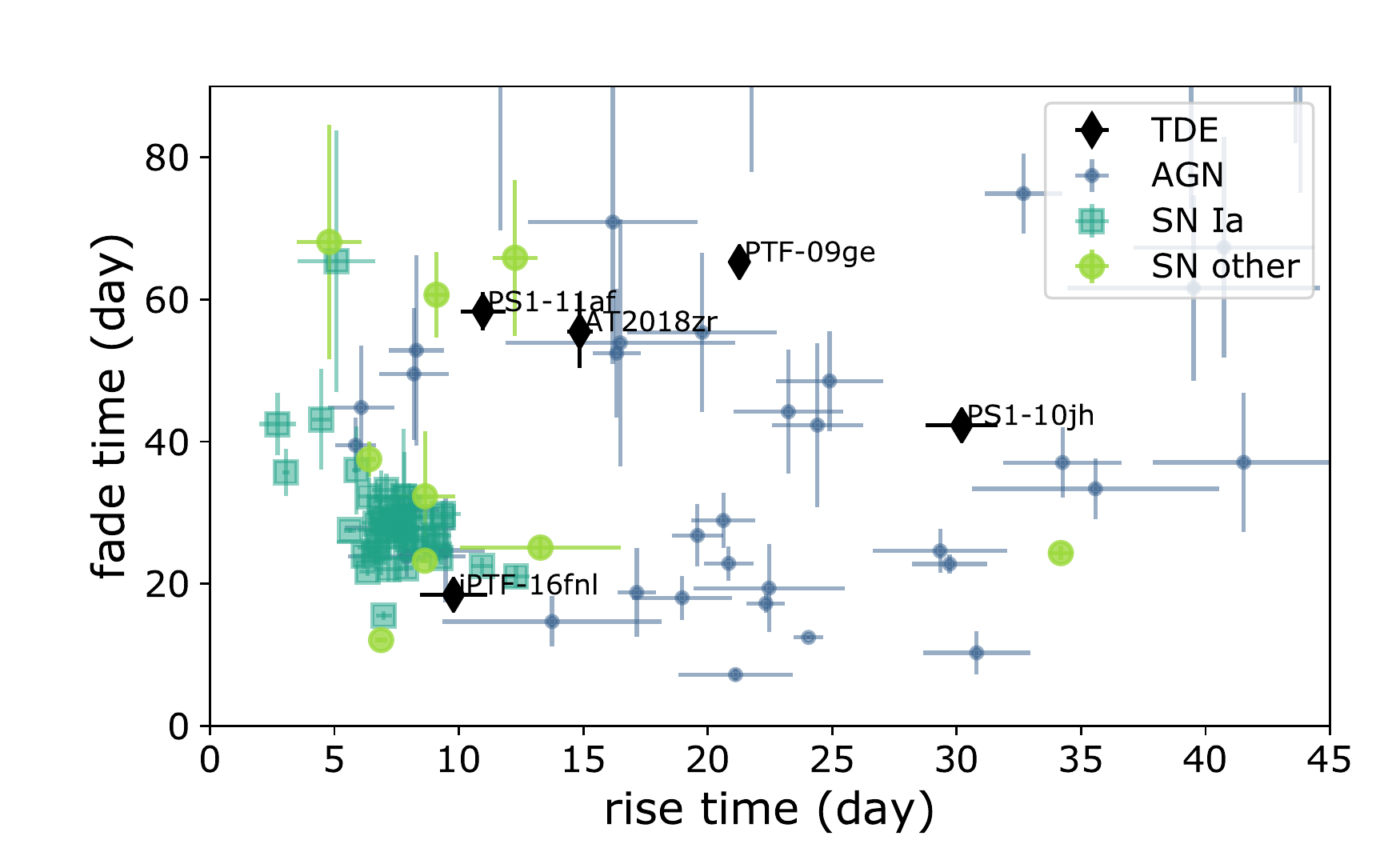}\\
\includegraphics[width=0.45 \textwidth, trim=3mm 3mm 4mm 6mm, clip]{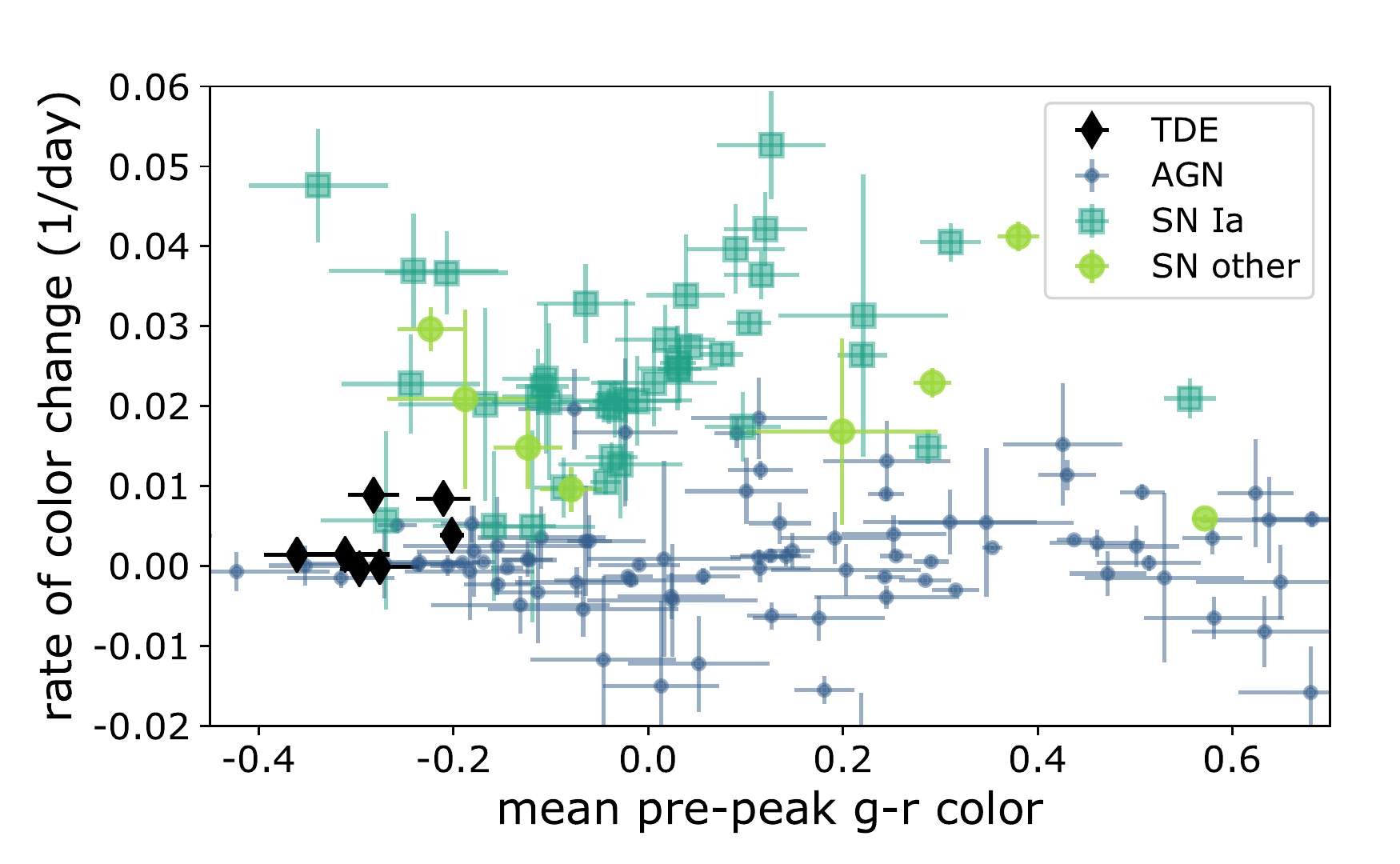}
\caption{Tidal disruption flares compared to other nuclear flares and transients detected by ZTF. {\it Top:} Rise timescale versus the fade timescale,  both measured using the $g$/$r$ band observations, see Eq.~\ref{eq:lcmodel}}. Known TDEs have a longer rise and fade timescale compared to most SNe. {\it Bottom:} the mean $g$-$r$ color versus its slope (Eq.~\ref{eq:color}) as measured during the first 100 days since the peak of the light curve. Unlike known SNe, all known TDEs have a near-constant color. Most AGN flares also have a constant color, but their mean colors (as measured in the difference image) show a much larger dispersion. \label{fig:compare}
\end{figure}

\section{Comparison to known TDEs, SNe, and AGN flares}\label{sec:compare}
To date, only four\footnote{PTF-09djl and PTF-09axc \citep{Arcavi14} also have pre-peak detections, but no post-peak detections; ASASSN-18pg/AT2018dyb is reported \citep{Brimacombe18} to be detected on the rise, but its light curve has not been published yet.} published optical TDEs have a resolved rise-to-peak: PS1-10jh \citep{Gezari12}, PS1-11af \citep{Chornock14}, PTF-09ge \citep{Arcavi14} and iPTF-16fnl \citep{Blagorodnova17}. The earliest ZTF detection of AT2018zr is 50~days before the peak of the light curve. Measurements of the rise time to peak are important because this parameter is expected to scale with the mass of the black hole that disrupted the star \citep{Rees88,Lodato09}. In Fig.~\ref{fig:lcs} we show the light curves of the five TDEs with pre-peak detections. We compare both the rest-frame $g$-band luminosity and the blackbody luminosity. The $k$-correction and bolometric correction were estimated using the mean blackbody temperature of the post-peak observations \citep{vanVelzen18_FUV}, except for AT2018zr, for which we used the temperature estimated from the nearest {\it Swift}/UVOT observation. 

Our sample of nuclear flares from ZTF data is large enough to provide a meaningful comparison of the photometric properties of TDEs, SNe, and AGN flares. Selecting alerts that were discovered between May 1 and August 8, we obtain 840 sources. Of these, 331 can be classified as AGN using  the Million quasar catalog \citep[][v5.2.]{Flesch15}. Our sample contains 81 spectroscopically confirmed SNe (of which 62 are SNe Type Ia) and 3 cataclysmic variables (CVs). The spectroscopic observations for SN/CV classifications were obtained by the ZTF collaboration, with SEDM serving as the main instrument; SNe typing was established using SNID \citep{Blondin07}.

To be able to compare TDEs, SNe, CVs, and AGN flares, we use a single light-curve model to describe the observations of these transients. We compute the fading timescale of the light curve with respect to the rise time to the peak of the light curve using an exponential decay, while rise-to-peak is modeled using a Gaussian function:
\begin{equation}\label{eq:lcmodel}
    F(t) = F_{\rm peak} \times 
    \begin{cases}  e^{-(t-t_{\rm peak})^2/2\sigma^2)} & t\leq t_{\rm peak} \\ 
    e^{-(t-t_{\rm peak})/\tau} & t>t_{\rm peak}\\
    \end{cases}
\end{equation} 
This simplistic light-curve model is a compromise between using specific models for each type of object (e.g., SN Ia templates) and using a completely non-parametric approach (e.g., interpolating the light curve to measure the FWHM). Using all available ZTF data (i.e., including upper limits prior to the first detection of the alert), we fit both the $r$-band and $g$-band simultaneously. To model the SED, we use a constant color for the observation before the peak of the light curve, and a linear change of color with time for the post-peak observations:
\begin{equation}\label{eq:color}
    {(g-r)}(t) = \begin{cases} \left<(g-r)\right>& t\leq t_{\rm peak} \\  \left<(g-r)\right>+ a\,(t-t_{\rm peak}) & t>t_{\rm peak}\\
    \end{cases}
\end{equation}
The use of a constant color before the peak helps to get robust results from the fitting procedure (pre-peak light curves often contain too few points to constrain any color evolution), plus this light-curve model also matches the observed behavior of SN Ia, which only show significant cooling after maximum light \citep[e.g.,][]{Hoeflich17}.

To summarize, our light-curve model has six free parameters: rise timescale ($\sigma$),  fade timescale ($\tau$), the time of peak ($t_{\rm peak}$), flux at peak ($F_{\rm peak}$), mean pre-peak color ($\left<g-r\right>|_{t<t_{\rm peak}}$), and rate of color change ($a$, with units time$^{-1}$).

For AT2018zr, we have no ZTF photometry post peak (because the flux of the transient is contained in the reference image of the public survey, which is not available for reprocessing until the first ZTF data release; \citealt{Graham19}) and we instead use the {\it Swift}/UVOT photometry. For the other four TDEs with a resolved peak of the light curve, we include the published photometry\footnote{Obtained using the Open TDE Catalog, \url{http://TDE.space}. } up to 100~days post-peak (when a longer temporal baseline is used, an exponential decay no longer provides a good description of the TDE light curve). 

In Fig.~\ref{fig:compare} we show the result of applying our light-curve model to AT2018zr, other known TDEs, as well as the AGN, SNe and CVs in our sample of nuclear flares. We discuss these results in Section~\ref{sec:dics_compare}. 

\begin{figure}
\centering
\includegraphics[width=0.45 \textwidth, trim=3mm 3mm 4mm 6mm, clip]{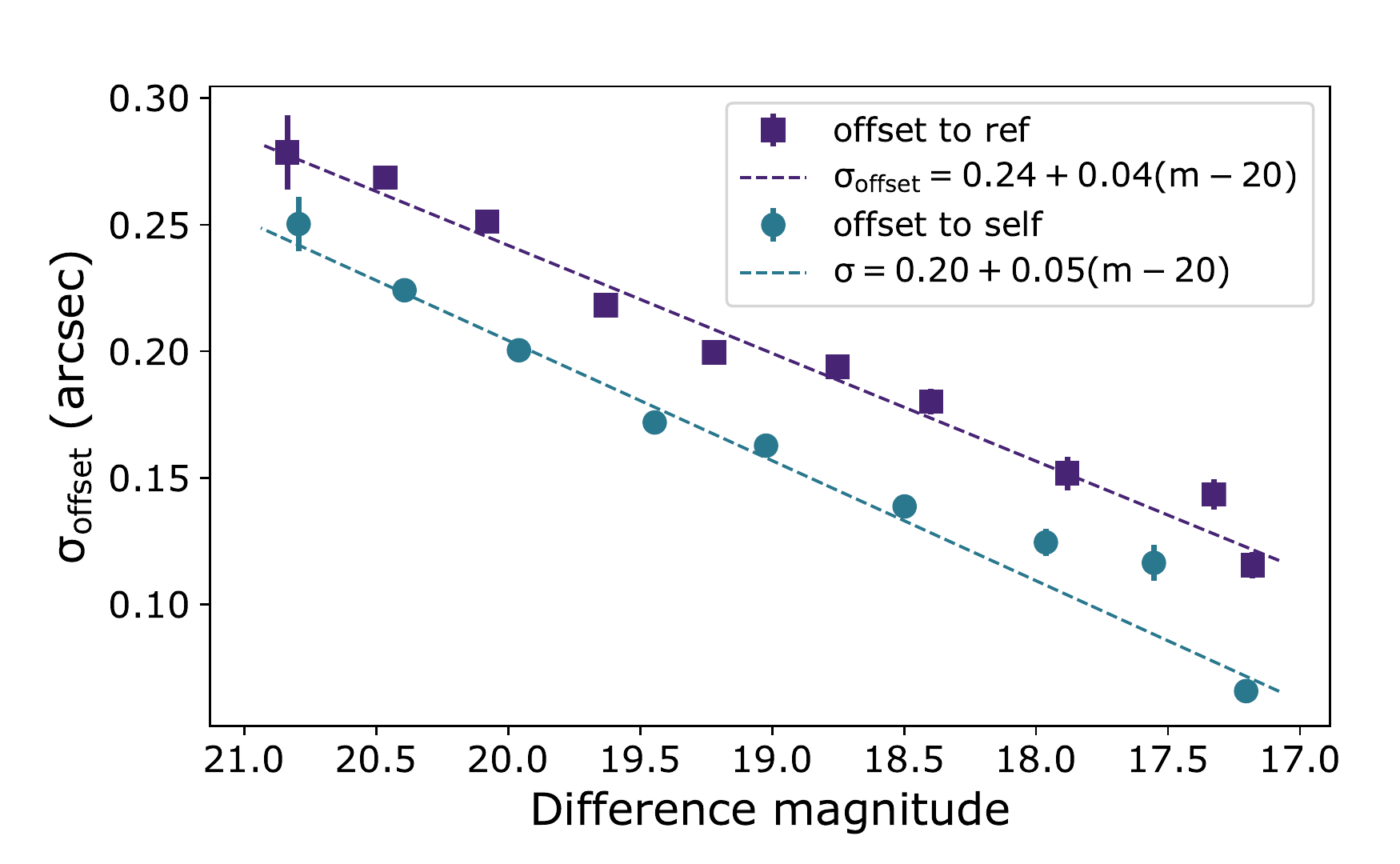} \qquad
\caption{ZTF astrometric accuracy for nuclear flares. The rms of the angular offset of AGN flares as a function of the magnitude in the difference image is shown. We show both the offset to the centroid of the reference image (squares) and the angular distance to the median location of the source in the difference images (circles). We used 11476 offset measurements (both R.A. and Decl.) of 128 AGN. 
}\label{fig:astro_unc}
\end{figure}

\section{Host-flare astrometry in ZTF data}\label{sec:offset}
In the previous section we found that our sample of nuclear flares contains about 10\% spectroscopically confirmed SNe. As explained in Section~\ref{sec:nuc_select}, the sample of nuclear flares was constructed from alerts with at least one detection with a host-flare distance smaller than 0\farcs6. However we expect that the mean host-flare distance can be measured with a precision that is better than 0\farcs6, thus facilitating a better separation of nuclear flares (AGN/TDEs) and SNe. 

To understand how the measurement of the offset scales with the signal-to-noise ratio of the detection, we collected ZTF measurements for a sample of known AGN. To obtain a good measurement of the mean and rms of the offset, we required at least seven detections and a median host-flare distance $<0\farcs3$, leaving 128 AGN. Under the assumption that the variability of these sources originates from the photometric center of their host galaxy, the observed rms of the offset yields the uncertainty, $\sigma_{\rm offset}$. 

\begin{figure}
\includegraphics[width=0.45 \textwidth, trim=3mm 3mm 4mm 6mm, clip]{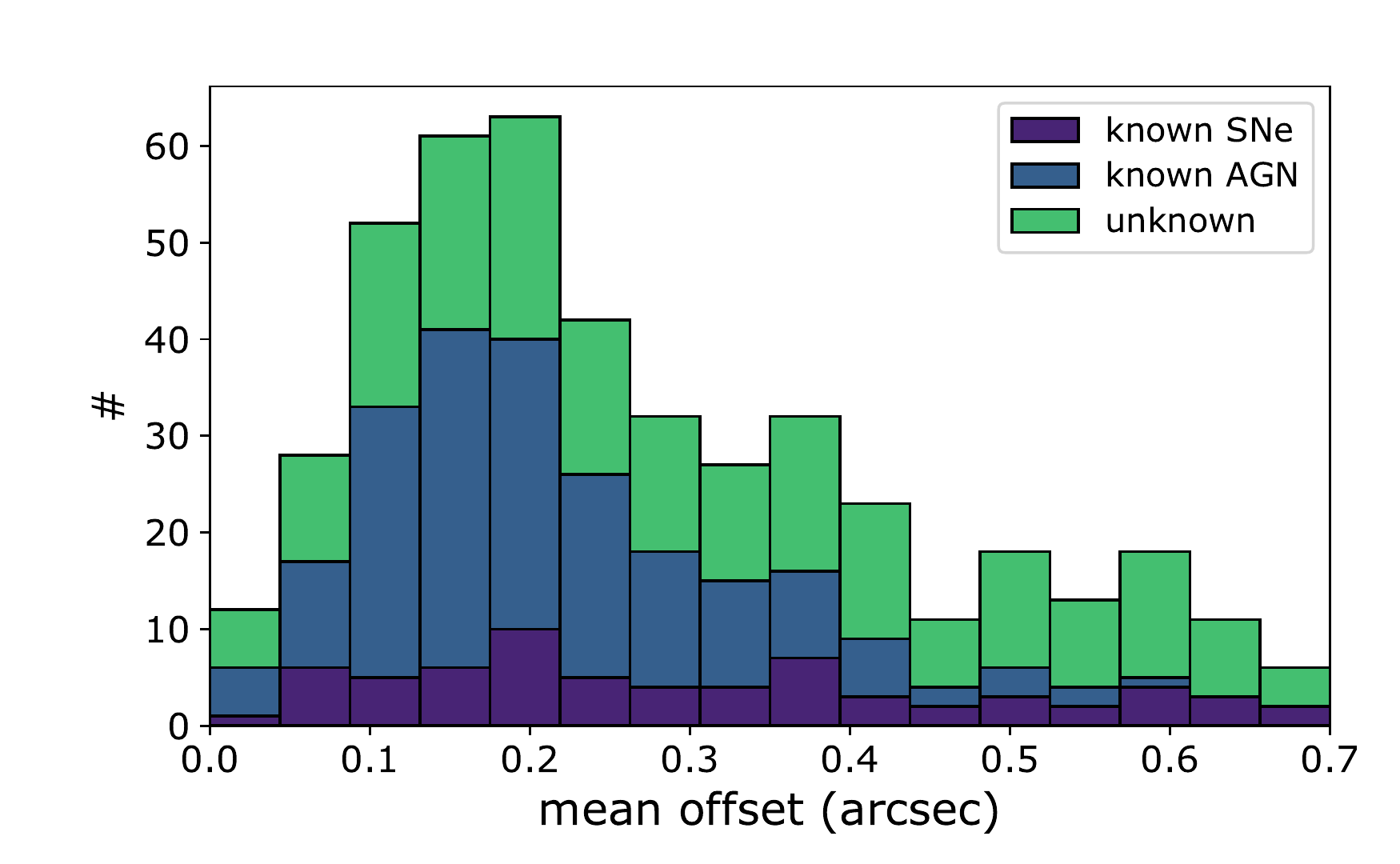}
\caption{Stacked histogram of the weighted (Eq.~\ref{eq:astro_unc}) mean host-flare distance for sources in our sample of nuclear flares, selected from 3 months of ZTF observations. }\label{fig:astro_histo}
\end{figure}

In Fig.~\ref{fig:astro_unc} we show $\sigma_{\rm offset}$ binned by the PSF magnitude of the flare in the difference image ($m_{\rm diff}$). 
We find the following dependence between these two parameters 
\begin{equation}\label{eq:astro_unc}
    \sigma_{\rm offset} = 0.24+ 0.04(m_{\rm diff}-20)~{\rm arcsec.} 
\end{equation}
We can use this relation to compute the inverse-variance weighted mean of the offset. In Fig.~\ref{fig:astro_histo} we show the weighted mean host-flare distance for the SNe, AGN, and unclassified flares in our nuclear flare sample (again using only sources with at least seven detections). Multiple observations of the same flare lead to an increased accuracy for the mean host-flare distance, yielding a typical uncertainty of 0$\farcs$2 for nuclear flares with a few tens of detections (see the peak of the AGN distribution in Fig.~\ref{fig:astro_histo}). 

The measurements of the host-flare offset are not independent because each measurement depends on the same reference frame to yield the position of the host. To show the astrometric accuracy without the contribution of the reference frame, we also report the rms of the offset with respect to the median position of the flare in the difference image (Fig.~\ref{fig:astro_unc}). 
%For sources with a large number of detections, or for faint host galaxies, the uncertain on the mean offset is dominated by the contribution of the reference frame. 

\section{Discussion}\label{sec:dics}

\subsection{Origin of the thermal emission mechanism}\label{sec:disc_emission}
\begin{figure}
\centering
\includegraphics[width=0.45\textwidth, trim=3mm 3mm 4mm 6mm]{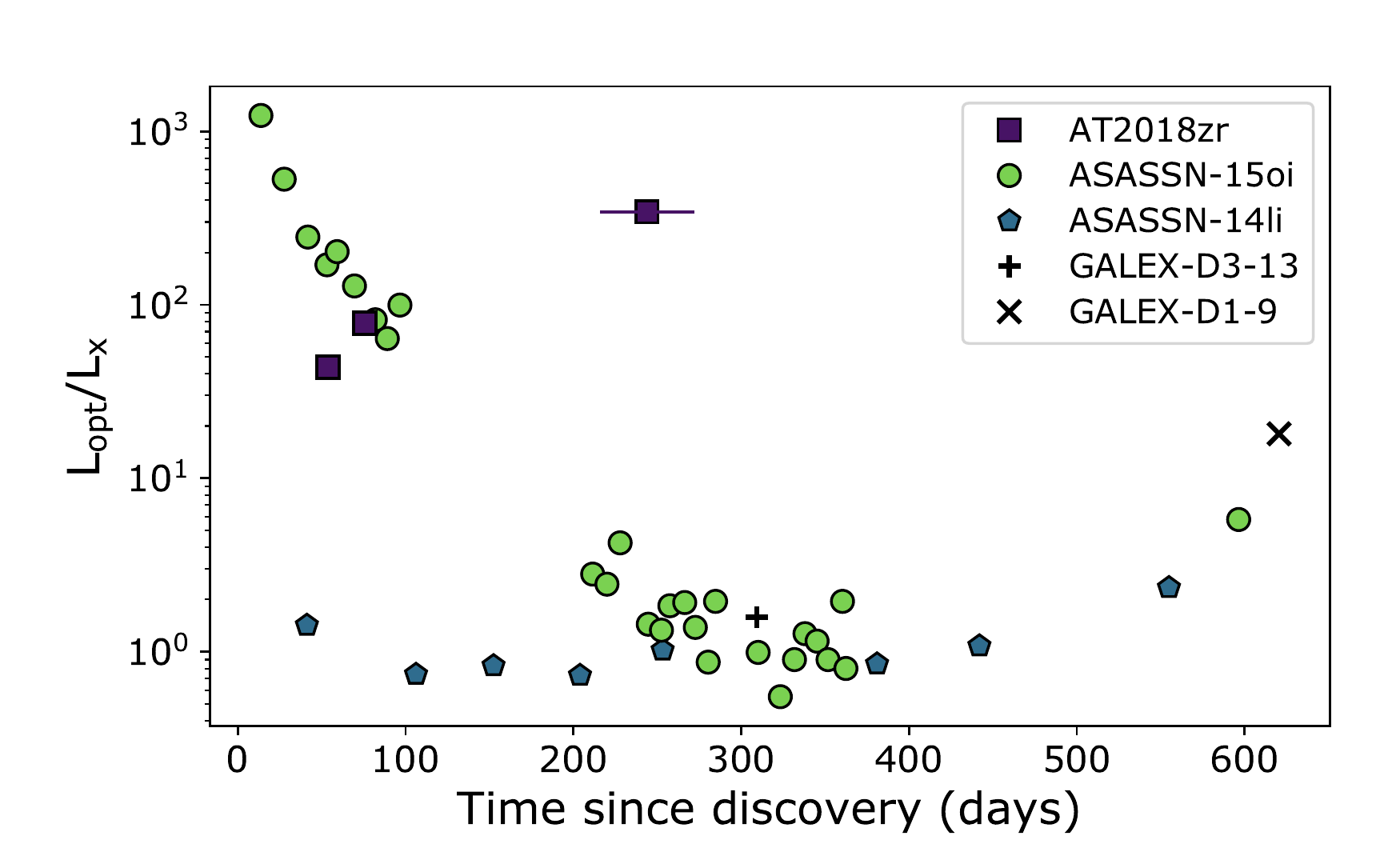} 
\caption{The ratio of the optical/UV blackbody luminosity to the X-ray luminosity (0.3-10 keV) as as a function of time. For AT2018zr, the first two points are based on XMM observations, while the third point is based on binned XRT observations (see Table~\ref{tab:xmm}). Data on previous TDEs taken from \citet{Gezari17a}.}\label{fig:ratio}
\end{figure}

AT2018zr is the fifth optical/UV-selected TDE with an X-ray detection, and only the third source that was detected within a few months of its discovery. Close to peak, its optical-to-X-ray flux ratio is similar to that of ASASSN-15oi (Fig.~\ref{fig:ratio}). However, contrary to previous TDEs, AT2018zr shows an increase of $L_{\rm opt}/L_X$; its late-time X-ray luminosity remained approximately constant (Table~\ref{tab:xmm}), while the UV/optical blackbody luminosity increased.

The low X-ray luminosity of AT2018zr could be explained by a delay in the formation of the accretion disk, similar to the explanation proposed for the X-ray behavior of ASASSN-15oi \citep{Gezari17a}. Yet, contrary to this TDE, the X-ray luminosity of AT2018zr remains weak compared to the optical/UV luminosity, suggesting no significant disk has formed yet.
If instead disk formation is efficient and the optical emission is explained by reprocessing of emission from the inner disk, the low X-ray luminosity of AT2018zr must be explained by obscuration of this disk. 
However the increase of $L_{\rm opt}/L_X$ presents a challenge to this explanation. We generally expect that the optical depth for X-ray photons decreases with time because the processing layer expands/dilutes \citep[e.g.,][]{Metzger16} and this would yield a {\it decrease} of  $L_{\rm opt}/L_{X}$. 

The blackbody radius corresponding to the single temperature model of the X-ray spectrum is $\approx 2\times 10^{10}$~cm for the first epoch of X-ray observations. This corresponds to the Schwarzschild radius ($R_s$) of a black hole with a mass of $5\times 10^4$~$M_\odot$. Such a low-mass black hole is not expected given the properties of the host galaxy of AT2018zr. If half of the stellar mass of the galaxy is in the bulge, the predicted black hole mass \citep{Gultekin09} is $8\times 10^6$~$M_\odot$ (consistent with the black hole mass estimate of \citealt{Holoien18a}). If the observed X-ray photons originated from the inner part of an accretion disk, the intrinsic X-ray luminosity must be $\sim 10^3$ times higher to match the blackbody radius to the expected size of the inner disk. Part of this tension can be alleviated if the observed X-ray temperature is higher than the true temperature due to obscuration or a contribution of inverse Compton emission to the 0.3-1~keV spectrum.  For example, if we adopt the upper limit to the intrinsic absorption from the X-ray spectral fit of $N_{\rm H} = 8 \times 10^{21} {\rm cm}^{-2}$, then the inferred blackbody temperature decreases by a factor of 2 and the unabsorbed X-ray luminosity increases by a factor of 1800, yielding a blackbody radius that is a factor of 170 larger and of the same of order as the expected size of the inner disk.

We thus conclude that if a modest amount of neutral gas has affected the intrinsic X-ray spectrum, we can explain the unexpectedly small radius of the X-ray photosphere inferred from the observed spectrum.  Based on the increase of $L_{\rm opt}/L_{X}$, we suggest this obscuring material would be outside the optical photosphere, i.e., the absorbed X-ray energy is not the source of the observed optical/UV emission of AT2018zr.

Even after accounting for the potential effect of absorption on the X-ray spectrum, the X-ray blackbody radius from our {\em XMM-Newton} observations is two orders of magnitude smaller than the inner radius of  $\sim 500 R_s$ of the elliptical disk model proposed by \citet{Holoien18a}. While this extended elliptical disk could explain the properties of the optical emission lines, our X-ray observations suggest a small, compact accretion disk is present as well.  

Based on {\it Swift}/UVOT observations that covered the first three months after peak, \citet{Holoien18a} reported a modest (40\%) increase in the optical/UV blackbody temperature. After a gap in the {\it Swift} coverage (due to Sun constraints), we find that the temperature has increased by at least a factor of 3; $T \gtrsim 5 \times 10^4$~K in the latest {\it Swift} observations, obtained 250~days after peak. The result of this temperature evolution is an flattening of the UV light curve (Fig.~\ref{fig:lc}) and an increase of the blackbody luminosity (Fig.~\ref{fig:lcs}). The UV/optical blackbody radius has decreased by an order of magnitude: from $10^{15.1}$~cm for the observations near peak to $\lesssim 10^{14}$~cm. We note that the {\it Swift} UV observations near peak show a hint of a blue excess (Fig.~\ref{fig:sed}). This suggests the hot component detected at late-time was also present at early-time and became more prominent due to the fading of the lower-temperature component. 

The observed flattening of the UV light curve of AT2018zr (Fig.~\ref{fig:lc}) is a common feature of TDEs and can be interpreted as an increased contribution of an accretion disk to the SED \citep{vanVelzen18_FUV}. The current blackbody temperature of AT2018zr (about 250 days post-peak) is also similar to the temperature of most TDEs that have been detected at UV wavelengths 1--10 years after peak \citep{vanVelzen18_FUV}.
However, the factor 4 increase of the blackbody temperature of AT2018zr is uncommon \citep[cf.][their Fig. 11]{Hung17}. The light curve of the TDE ASASSN-15oi \citep{Holoien16b} showed a temperature increase of a factor $2$ (and a factor of $10$ decrease of the blackbody radius) during the first 100 days of observations. We note that the early-time optical/UV temperature of AT2018zr ($T=10^{4.15}$~K) was relatively low, which could explain why the temperature increase is larger compared to most previous TDES. The TDE PTF-09ge also displayed a relatively low temperature near peak (as measured from the $g$-$r$ color), followed by a factor of 4 increase to the blackbody temperature inferred from HST observations obtained 6 years after peak \citep{vanVelzen18_FUV}.

\begin{figure}
\centering
\includegraphics[width=0.45 \textwidth, trim=3mm 3mm 4mm 6mm, clip]{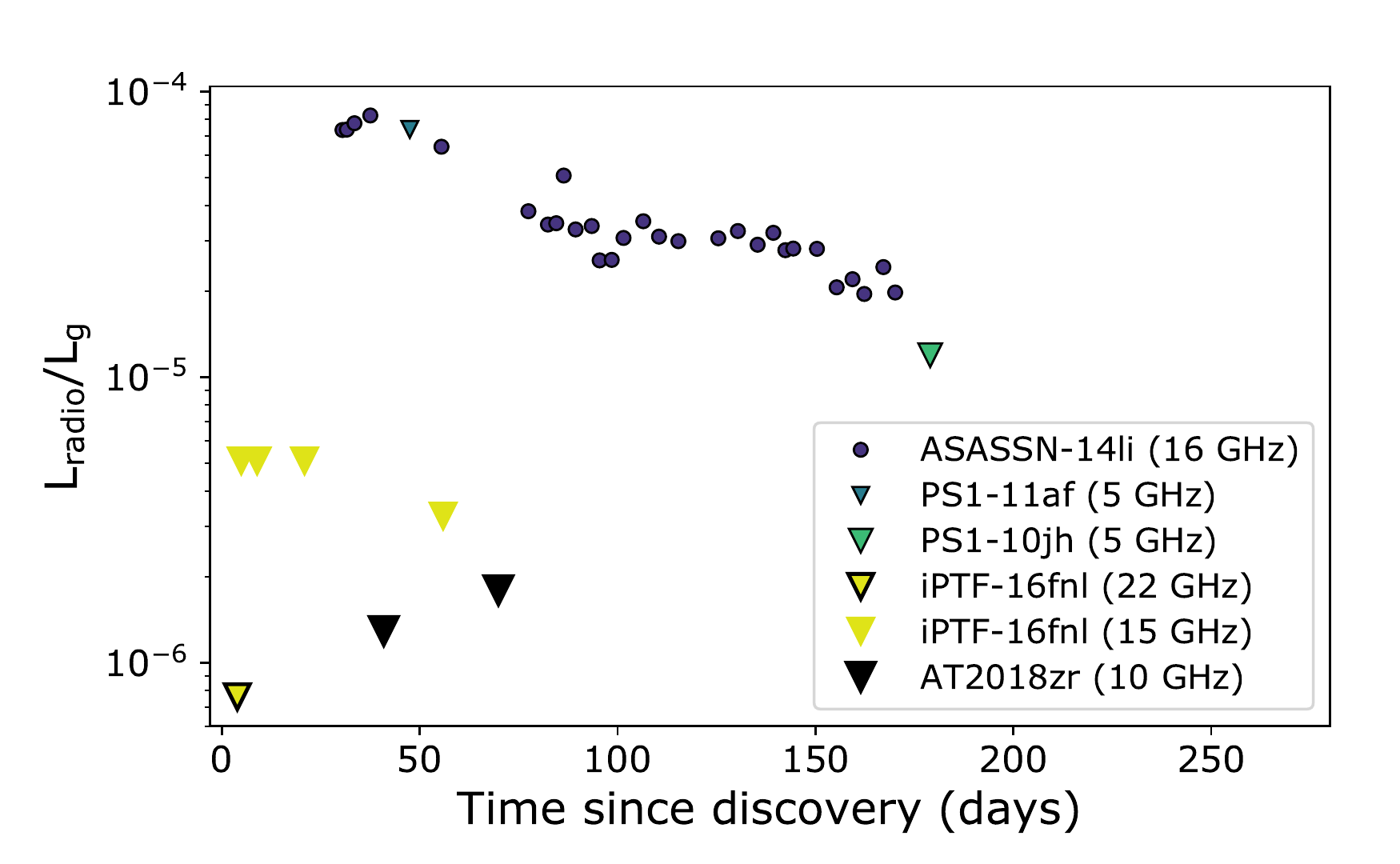}
\caption{Radio luminosity of optical TDEs, normalized to the peak of the optical light curves ($\nu L_\nu$ in the rest-frame $g$-band; see Fig.~\ref{fig:lcs}). Triangles indicate 3$\sigma$ upper limits to the radio luminosity. We only show TDEs with radio follow-up observations obtained within one year of the first optical detection: ASASSN-14li \citep{vanVelzen16,Alexander16}, iPTF-16fnl \citep{Blagorodnova17}, PS1-11af \citep{Chornock14}, PS1-10jh \citep{Gezari12,vanVelzen12b}.}
\label{fig:radio}
\end{figure}

\subsection{Interpretation of radio non-detection}
The upper limit to the radio luminosity of AT2018zr is one order of magnitude lower than the observed radio emission of the TDE ASASSN-14li \citep{vanVelzen16b,Alexander16,Bright2018}. Currently, only the TDE iPTF-16fnl \citep{Blagorodnova17} has received radio follow-up observations close to the peak of the flare with a similar sensitivity. This source was also not detected at radio frequencies. However, this flare was exceptional, being the faintest and fastest fading TDE to date (see Fig.~\ref{fig:lcs}). The optical properties of AT2018zr, on the other hand, are similar to the mean properties of the current TDE sample (see Figs.~\ref{fig:lcs} \& \ref{fig:compare} and \citealt{Hung17}).% including ASASSN-14li, which had a peak luminosity of $\sim 10^42.6\,{\rm erg}\,{\rm}^{-1}$ in the $g$-band \citep{vanVelzen18}.

Our radio non-detection rules out the hypothesis that TDEs with a typical optical luminosity and fade timescale produce radio emission similar to that of ASASSN-14li (Fig.~\ref{fig:radio}). However, the X-ray luminosity of AT2018zr is two orders of magnitude lower than ASASSN-14li \citep{Holoien16}. If the radio luminosity scales linearly with the power of the accretion disk, as observed for ensembles of radio-loud quasars \citep{Rawlings91,Falcke99,vanVelzen15} and for ASASSN-14li \citep{PashamvanVelzen17}, the expected radio flux of AT2018zr would be too faint to be detectable. 

Free--free absorption is unlikely to affect the 10~GHz flux of AT2018zr because it would require an unrealistically high electron density. For an electron temperature of $T_e=10^4$~K, we require an emission measure (EM) of at least $\sim 10^8$~pc\,cm$^{-6}$ to yield a significant optical depth ($\tau >1$) for free-free absorption at 10~GHz \citep[e.g.,][]{Condon92}. This limit on the EM corresponds to a mean electron density of $10^4$~cm$^{-3}$ within one parsec, which is at least two orders of magnitude larger than the particle density within the central parsec of our Galactic Center \citep{Baganoff03,Quataert04} or the circumnuclear density inferred from the jetted TDEs \citep{Berger12,Generozov17,Eftekhari18}. For higher electron temperatures, as expected for galaxy centers	\citep[][]{Lazio99}, the lower limit on the EM would increase even further. 

\subsection{Rise timescale and Black Hole Mass}
\begin{figure}
\centering
\includegraphics[width=0.45 \textwidth, trim=3mm 3mm 4mm 6mm, clip]{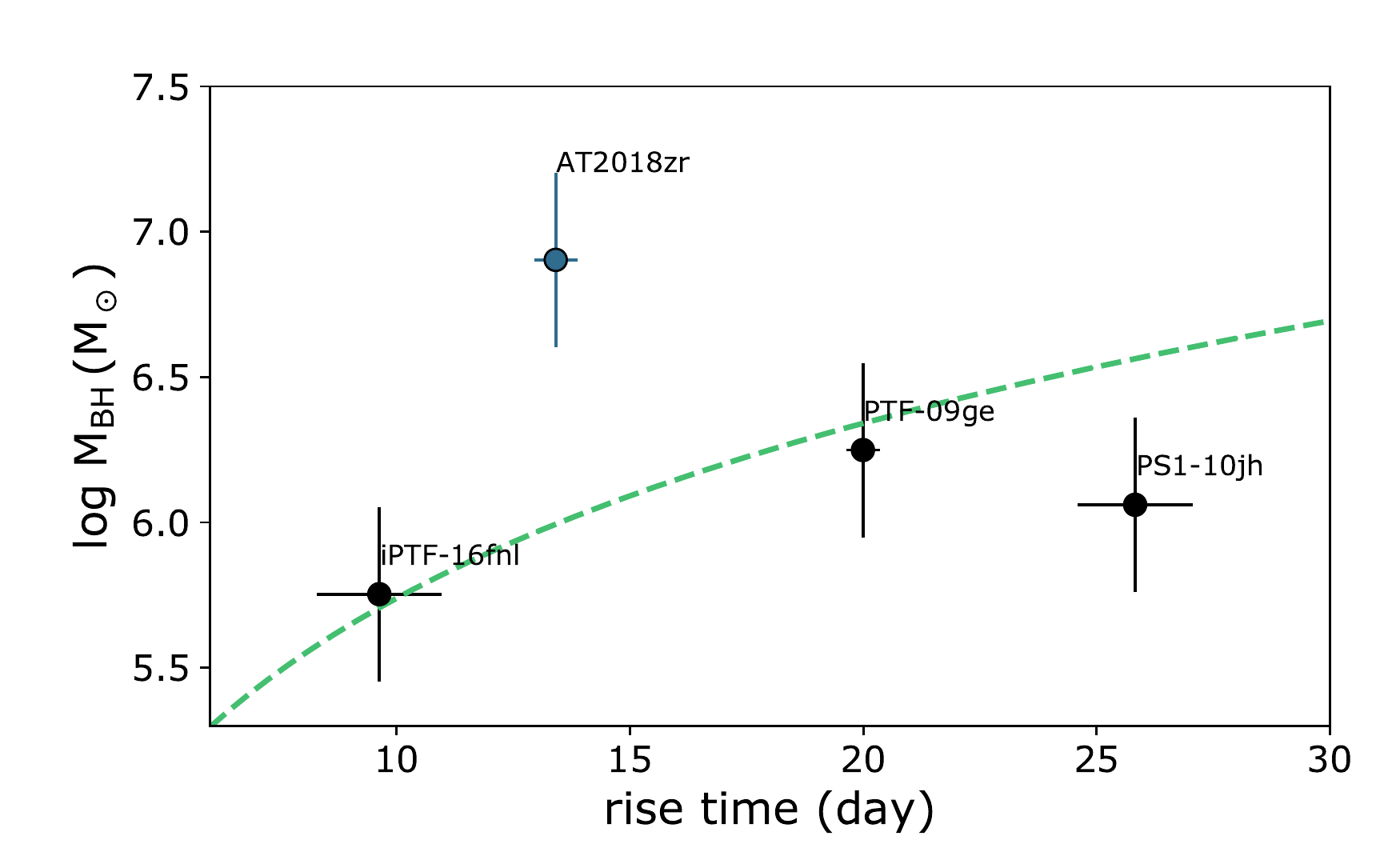}
\includegraphics[width=0.45 \textwidth, trim=3mm 3mm 4mm 6mm, clip]{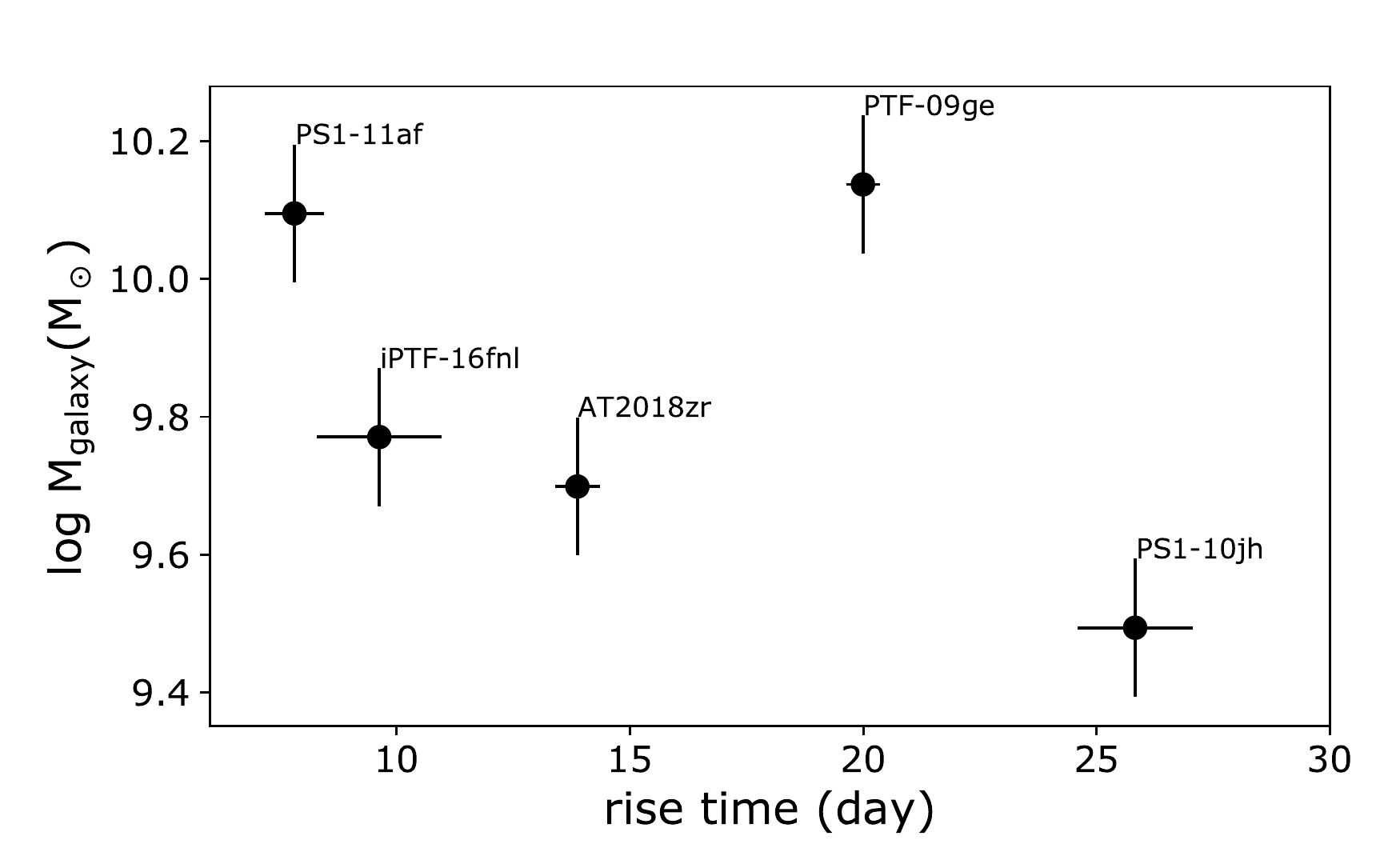}
\caption{The rest-frame rise timescale of the light curves, measured using a Gaussian function (Eq.~\ref{eq:lcmodel}), and the black hole mass ({\it top}) or the total stellar mass of the host galaxy ({\it bottom}). The black hole mass is estimated using the velocity dispersion of the host galaxy, with the exception of AT2018zr. The dashed line shows the expected scaling (Eq.~\ref{eq:tfb}) between the rise time to peak and black hole mass. }
\label{fig:MBH_rise}
\end{figure}

While our current sample of TDEs with a resolved rise-to-peak is still small, we can start to search for the anticipated correlation between black hole mass and rise timescale \citep{Rees88,Lodato09}. To estimate the black hole mass we use the velocity dispersion measurements from \citet{Wevers17} and the \citet{Gultekin09} $M$-$\sigma$ relation. For the host galaxy of AT2018zr, a velocity dispersion measurement is not yet available and we adopt the black hole mass from the bulge mass ($M_\bullet = 10^{6.9}\,M_\odot$, see Section~\ref{sec:disc_emission}). To provide a more uniform comparison we also consider the relation between rise time and total stellar mass. The results are shown in Fig.~\ref{fig:MBH_rise}. 

Our measurement of the rise time uses a Gaussian function (Eq.\ref{eq:lcmodel}), which has no defined start time. To be able to compare our measurement to the predicted fallback timescale ($t_{\rm fb}$) we assume the disruption happened at $t=3\sigma$ (i.e., when the flux in the model light curve is just 1\% of the flux at peak). In Fig.~\ref{fig:MBH_rise} we show the predicted rise time from the theoretical fallback time of \citet{Stone13}, for the disruption of a star with a mass of 1~$M_\odot$ and an impact parameter of unity,
\begin{equation}\label{eq:tfb}
    t_{\rm fb} =  3.5\times 10^6 \left(\frac{M_\bullet}{ 10^6~M_\odot}\right)^{1/2}~{\rm s} \quad.
\end{equation}

We find no correlation between the rise time and total galaxy mass. This could be considered surprising given that a correlation between the {\it fade} timescale and the black hole mass has been reported \citep{Blagorodnova17, Wevers17}.  However, our results also show that the rise and fade timescales themselves appear to be uncorrelated (Fig.~\ref{fig:compare}). It could be possible that the post-peak light curve provides a better tracer for the fallback rate---and thus a better mass estimate \citep{Mockler18}---compared to the rise-to-peak. 

\subsection{Photometric selection of TDEs in optical surveys}\label{sec:dics_compare}
Using only 3 months of ZTF data, we confirm the conclusions from earlier TDE population studies \citep{Gezari09,vanVelzen10,Holoien16b,Hung17}, showing that TDEs are a class of flares with a shared set of photometric properties. 

Our work is the first to quantify the distribution of rise and fade timescales of TDEs. We find that most TDEs have both a longer rise time and fade time compared to SNe (Fig.~\ref{fig:compare}, top panel). The TDE iPTF-16fnl is an interesting exception, displaying a rise timescale at the edge of the SN Ia distribution and a fade timescale that is {\it faster} compared to most SNe. 

Using only pre-peak observations, effective removal of SNe~Ia is possible by restricting to flares with a rise time ($\sigma$, Eq.~\ref{eq:lcmodel}) that is longer than $10$~days. While AGN flares can have a wide range of rise/fade timescales, very few rise and fade within a few months. Only 5\% of AGN in our sample of nuclear flares have rise and fade timescales that fall within the range spanned by known TDEs.

The largest contrast between TDEs and SNe is found when we consider the mean color and color change (Fig.~\ref{fig:compare}). We see that TDEs cluster in the region of blue and constant colors \citep{vanVelzen10}. Near their peak, some SNe can be as blue as TDEs, yet these SNe also cool very fast. 

%We find no correlation between the rise timescale and fade timescale of  TDEs. Since the rise timescale is expected to contain most information about the mass of the black hole that disrupted the star, this casts some doubt on the claim that this  mass can be measured using only the fading part of the light curve \citep{Mockler18}. 

In this work we have demonstrated how several photometric properties can be used to separate TDEs from AGN flares and SNe: the rise/fade timescale (Fig.~\ref{fig:compare}, top panel), the flare's color and its evolution (Fig.~\ref{fig:compare}, bottom panel), and the location of the flare in the host galaxy (Fig.~\ref{fig:astro_histo}). Additional selection on the host galaxy properties can be used to further reduce AGN contamination (e.g., prior variability, the amplitude of the flux increase). While each of these metrics has exceptions (e.g., TDE from faint galaxies have large astrometric uncertainties on the host-flare offset, and some TDE rise and fade rapidly), photometric selection will be unavoidable in the era of the Large Synoptic Survey Telescope (LSST). The TDEs detected by LSST will be too faint and too numerous \citep[$\sim 1000$ per year;][]{vanVelzen10} to use spectroscopic follow-up observations for classification.

\acknowledgments
{\bf Acknowledgments}  ---
We thank the referee for the useful comments. 

This work was based on observations obtained with the Samuel Oschin Telescope 48-inch and the 60-inch Telescope at the Palomar Observatory as part of the Zwicky Transient Facility project. Major funding has been provided by the U.S National Science Foundation under Grant No. AST-1440341 and by the ZTF partner institutions: the California Institute of Technology, the Oskar Klein Centre, the Weizmann Institute of Science, the University of Maryland, the University of Washington, Deutsches Elektronen-Synchrotron, the University of Wisconsin-Milwaukee, and the TANGO Program of the University System of Taiwan.

We thank the National Radio Astronomy Observatory (NRAO) staff for the rapid scheduling of the VLA observations. NRAO is a facility of the National Science Foundation operated under cooperative agreement by Associated Universities, Inc. We thank the staff of the Mullard Radio Astronomy Observatory for their assistance in the operation of AMI. We acknowledge the use of public data from the Swift data archive. 

This research made use of Astropy, a community-developed core Python package for Astronomy \citep{Astropy-Collaboration18}.

S. Gezari is supported in part by NSF CAREER grant 1454816 and NSF AAG grant 1616566. M. M. Kasliwal acknowledges support by the GROWTH (Global Relay of Observatories Watching Transients Happen) project funded by the National Science Foundation PIRE (Partnership in International Research and Education) program under Grant No 1545949. N.R. acknowledges the support of a Joint Space-Science Institute prize postdoctoral
fellowship. J.C.A.M.-J. is supported by an Australian Research Council Future Fellowship (FT140101082). This project has received funding from the European Research Council (ERC) under the European Union's Horizon 2020 research and innovation programme (grant agreement No. 759194 - USNAC).

\software{Astropy \citep{Astropy-Collaboration18}, 
CASA \citep{McMullin07}, 
HEAsoft \citep{Arnaud96},
SAS \citep{Gabriel04},
FSPS \citep{Conroy09,Conroy10} with Python binding from dfm (\url{http://dx.doi.org/10.5281/zenodo.12157})
}

\bibstyle{aasjournals}

% if you have references to add that are not in sjoert.bib, put them in others.bib
\bibliography{general_desk,others}

\begin{deluxetable}{c c}
\tablewidth{\textwidth}
\tablecolumns{2}
\tablecaption{Synthetic host magnitudes}
\tablehead{~~~~~Filter~~~~~ & ~~~~~Magnitude~~~~~ }
\startdata
V      & 18.49 \\
B      & 19.40 \\
U      & 20.81 \\
UVW1   & 22.48 \\
UVM2   & 23.61 \\
UVW2   & 23.91 \\
\enddata
\tablecomments{Obtained by convolving the best-fit galaxy model (Fig.~\ref{fig:sed}) with the {\it Swift}/UVOT filter throughput. Not corrected for Galactic extinction. }
\label{tab:host_syn}
\end{deluxetable}

\begin{deluxetable}{l c c c c c c}
\tablewidth{\textwidth}
\tablecolumns{6}
\tablecaption{X-ray observations}
\tablehead{Start & Instrument 	& Int. time\tablenotemark{a} 			& Counts 	& $N_H$ 		& $T$ 	& Flux\tablenotemark{b}\\ 
		(MJD)	&	&(ks)		&			& (cm$^{-2}$)	& (eV) & ($\times 10^{-14}$ erg\,s$^{-1}$\,cm$^{-2}$)}
\startdata
58219.98		& XMM		& 20		&	259		& $<8\times 10^{21}$		& $97^{+17}_{-14}$	& $2.0^{+0.2}_{-0.5}$\\
58241.92		& XMM		& 25		&	190		& $<8\times 10^{21}$		& $114^{+21}_{-16}$	& $1.1^{+0.1}_{-0.1}$\\
58381.59--58438.77\tablenotemark{c}	& XRT			& 77		&	14		& --		& $100$\tablenotemark{d}	& $0.9\pm 0.6$\\
\enddata
\tablenotetext{a}{The time on-source.}
\tablenotetext{b}{Flux at 0.3-1~keV.}
\tablenotetext{c}{Binned XRT observations in this MJD range.}
\tablenotetext{d}{Value fixed.}
\tablecomments{Errors correspond to a 90\% confidence level.}
\label{tab:xmm}
\end{deluxetable}

\begin{deluxetable}{l l c c c}
\tablecolumns{4}

\tablecaption{Radio observations}
\tablehead{Instrument & Start  	&Int.\tablenotemark{a} 	&  rms & Flux$^b$ 	\\ 
			(MJD)	    &	&(min)		&($\mu$Jy/beam)	&($\mu$Jy)}
\startdata
AMI (16~GHz)        &  58205.8			    & 237.6	 	&   40.0 & $<120$\\
VLA X-band (10~GHz) & 58207.16				& 6.0	& 	9.1 & $<27$\\
VLA X-band (10~GHz) & 58236.14				& 6.1	& 	12.5 & $<37.5$ \\
\enddata
\tablenotetext{a}{The time on source.}
\tablenotetext{b}{The 3$\sigma$ upper limit to the flux.}\label{tab:janksy}
\end{deluxetable}

\startlongtable
\begin{deluxetable}{l c c c}

\tablecaption{Optical/UV photometry}
\tablewidth{0pt}
\tablecolumns{4}
\tabletypesize{\footnotesize}
\tablehead{MJD & Instrument & Filter & Mag}
\startdata
58099.480 & ZTF/P48       & r    & $>21.82$ \\ 
58100.470 & ZTF/P48       & r    & $>21.36$ \\ 
58101.360 & ZTF/P48       & r    & $>21.44$ \\ 
58102.440 & ZTF/P48       & r    & $>21.03$ \\ 
58104.440 & ZTF/P48       & r    & $>21.48$ \\ 
58105.310 & ZTF/P48       & r    & $>21.13$ \\ 
58154.220 & ZTF/P48       & r    & $>21.65$ \\ 
58155.230 & ZTF/P48       & r    & $>22.10$ \\ 
58156.230 & ZTF/P48       & r    & 21.06 $\pm$ 0.16 \\ 
58158.230 & ZTF/P48       & r    & 20.85 $\pm$ 0.12 \\ 
58160.230 & ZTF/P48       & r    & 20.47 $\pm$ 0.16 \\ 
58182.190 & ZTF/P48       & r    & 18.16 $\pm$ 0.07 \\ 
58183.180 & ZTF/P48       & r    & 17.98 $\pm$ 0.02 \\ 
58183.340 & SEDM/P60      & r    & 17.87 $\pm$ 0.13 \\ 
58190.154 & ZTF/P48       & r    & 17.59 $\pm$ 0.01 \\ 
58190.155 & ZTF/P48       & r    & 17.59 $\pm$ 0.02 \\ 
58222.136 & SEDM/P60      & r    & 18.23 $\pm$ 0.02 \\ 
58222.142 & SEDM/P60      & r    & 18.26 $\pm$ 0.02 \\ 
58222.148 & SEDM/P60      & r    & 18.25 $\pm$ 0.01 \\ 
58222.155 & SEDM/P60      & r    & 18.26 $\pm$ 0.01 \\ 
58222.161 & SEDM/P60      & r    & 18.24 $\pm$ 0.01 \\ 
58227.162 & SEDM/P60      & r    & 18.37 $\pm$ 0.03 \\ 
58227.168 & SEDM/P60      & r    & 18.40 $\pm$ 0.03 \\ 
58227.174 & SEDM/P60      & r    & 18.32 $\pm$ 0.03 \\ 
58204.533 & Swift/UVOT    & V    & 17.86 $\pm$ 0.30 \\ 
58213.031 & Swift/UVOT    & V    & 17.86 $\pm$ 0.42 \\ 
58237.078 & Swift/UVOT    & V    & 17.97 $\pm$ 0.36 \\ 
58129.330 & ZTF/P48       & g    & $>21.91$ \\ 
58131.340 & ZTF/P48       & g    & $>21.79$ \\ 
58132.360 & ZTF/P48       & g    & $>20.03$ \\ 
58166.260 & ZTF/P48       & g    & 18.85 $\pm$ 0.02 \\ 
58167.180 & ZTF/P48       & g    & 18.84 $\pm$ 0.03 \\ 
58168.164 & ZTF/P48       & g    & 18.77 $\pm$ 0.04 \\ 
58168.176 & ZTF/P48       & g    & 18.64 $\pm$ 0.03 \\ 
58170.392 & ZTF/P48       & g    & 18.64 $\pm$ 0.01 \\ 
58222.138 & SEDM/P60      & g    & 18.21 $\pm$ 0.03 \\ 
58222.144 & SEDM/P60      & g    & 18.20 $\pm$ 0.02 \\ 
58222.150 & SEDM/P60      & g    & 18.24 $\pm$ 0.02 \\ 
58222.157 & SEDM/P60      & g    & 18.25 $\pm$ 0.02 \\ 
58222.163 & SEDM/P60      & g    & 18.19 $\pm$ 0.02 \\ 
58227.164 & SEDM/P60      & g    & 18.29 $\pm$ 0.03 \\ 
58227.170 & SEDM/P60      & g    & 18.35 $\pm$ 0.02 \\ 
58204.527 & Swift/UVOT    & B    & 17.51 $\pm$ 0.12 \\ 
58208.840 & Swift/UVOT    & B    & 17.58 $\pm$ 0.22 \\ 
58210.375 & Swift/UVOT    & B    & 17.52 $\pm$ 0.19 \\ 
58212.637 & Swift/UVOT    & B    & 17.89 $\pm$ 0.26 \\ 
58213.029 & Swift/UVOT    & B    & 18.10 $\pm$ 0.28 \\ 
58215.023 & Swift/UVOT    & B    & 17.50 $\pm$ 0.21 \\ 
58221.212 & Swift/UVOT    & B    & 18.24 $\pm$ 0.22 \\ 
58223.266 & Swift/UVOT    & B    & 17.94 $\pm$ 0.21 \\ 
58228.507 & Swift/UVOT    & B    & 18.44 $\pm$ 0.44 \\ 
58231.961 & Swift/UVOT    & B    & 18.15 $\pm$ 0.34 \\ 
58204.526 & Swift/UVOT    & U    & 17.66 $\pm$ 0.09 \\ 
58208.839 & Swift/UVOT    & U    & 17.90 $\pm$ 0.17 \\ 
58210.375 & Swift/UVOT    & U    & 17.98 $\pm$ 0.17 \\ 
58212.637 & Swift/UVOT    & U    & 17.68 $\pm$ 0.15 \\ 
58213.028 & Swift/UVOT    & U    & 17.79 $\pm$ 0.14 \\ 
58215.023 & Swift/UVOT    & U    & 18.14 $\pm$ 0.20 \\ 
58221.212 & Swift/UVOT    & U    & 18.00 $\pm$ 0.12 \\ 
58223.266 & Swift/UVOT    & U    & 18.21 $\pm$ 0.16 \\ 
58228.507 & Swift/UVOT    & U    & 18.33 $\pm$ 0.23 \\ 
58231.961 & Swift/UVOT    & U    & 18.36 $\pm$ 0.23 \\ 
58234.018 & Swift/UVOT    & U    & 18.22 $\pm$ 0.18 \\ 
58237.072 & Swift/UVOT    & U    & 18.53 $\pm$ 0.17 \\ 
58240.131 & Swift/UVOT    & U    & 18.65 $\pm$ 0.16 \\ 
58242.317 & Swift/UVOT    & U    & 18.45 $\pm$ 0.32 \\ 
58249.364 & Swift/UVOT    & U    & 18.80 $\pm$ 0.20 \\ 
58252.145 & Swift/UVOT    & U    & 18.69 $\pm$ 0.23 \\ 
58255.732 & Swift/UVOT    & U    & 19.04 $\pm$ 0.34 \\ 
58261.116 & Swift/UVOT    & U    & 18.92 $\pm$ 0.26 \\ 
58264.034 & Swift/UVOT    & U    & 18.68 $\pm$ 0.22 \\ 
58381.594 & Swift/UVOT    & U    & 19.92 $\pm$ 0.48 \\ 
58410.805 & Swift/UVOT    & U    & 19.90 $\pm$ 0.38 \\ 
58417.716 & Swift/UVOT    & U    & 19.97 $\pm$ 0.40 \\ 
58438.768 & Swift/UVOT    & U    & 20.22 $\pm$ 0.47 \\ 
58204.525 & Swift/UVOT    & UVW1 & 17.89 $\pm$ 0.08 \\ 
58208.838 & Swift/UVOT    & UVW1 & 18.39 $\pm$ 0.14 \\ 
58210.374 & Swift/UVOT    & UVW1 & 18.12 $\pm$ 0.13 \\ 
58212.636 & Swift/UVOT    & UVW1 & 18.14 $\pm$ 0.13 \\ 
58213.027 & Swift/UVOT    & UVW1 & 18.12 $\pm$ 0.11 \\ 
58215.022 & Swift/UVOT    & UVW1 & 18.39 $\pm$ 0.15 \\ 
58221.210 & Swift/UVOT    & UVW1 & 18.51 $\pm$ 0.10 \\ 
58223.265 & Swift/UVOT    & UVW1 & 18.41 $\pm$ 0.12 \\ 
58228.506 & Swift/UVOT    & UVW1 & 18.40 $\pm$ 0.15 \\ 
58231.960 & Swift/UVOT    & UVW1 & 18.42 $\pm$ 0.15 \\ 
58234.017 & Swift/UVOT    & UVW1 & 18.63 $\pm$ 0.14 \\ 
58237.071 & Swift/UVOT    & UVW1 & 18.80 $\pm$ 0.12 \\ 
58240.129 & Swift/UVOT    & UVW1 & 18.79 $\pm$ 0.11 \\ 
58242.316 & Swift/UVOT    & UVW1 & 18.40 $\pm$ 0.19 \\ 
58249.362 & Swift/UVOT    & UVW1 & 18.68 $\pm$ 0.12 \\ 
58252.144 & Swift/UVOT    & UVW1 & 18.87 $\pm$ 0.14 \\ 
58255.731 & Swift/UVOT    & UVW1 & 18.59 $\pm$ 0.14 \\ 
58261.114 & Swift/UVOT    & UVW1 & 18.79 $\pm$ 0.13 \\ 
58264.032 & Swift/UVOT    & UVW1 & 18.69 $\pm$ 0.12 \\ 
58381.592 & Swift/UVOT    & UVW1 & 19.37 $\pm$ 0.16 \\ 
58396.660 & Swift/UVOT    & UVW1 & 19.73 $\pm$ 0.19 \\ 
58403.566 & Swift/UVOT    & UVW1 & 20.08 $\pm$ 0.44 \\ 
58410.803 & Swift/UVOT    & UVW1 & 20.00 $\pm$ 0.23 \\ 
58417.714 & Swift/UVOT    & UVW1 & 20.30 $\pm$ 0.29 \\ 
58431.138 & Swift/UVOT    & UVW1 & 20.14 $\pm$ 0.32 \\ 
58438.766 & Swift/UVOT    & UVW1 & 20.00 $\pm$ 0.23 \\ 
58204.537 & Swift/UVOT    & UVM2 & 18.06 $\pm$ 0.06 \\ 
58208.845 & Swift/UVOT    & UVM2 & 18.53 $\pm$ 0.09 \\ 
58210.379 & Swift/UVOT    & UVM2 & 18.38 $\pm$ 0.10 \\ 
58212.895 & Swift/UVOT    & UVM2 & 19.04 $\pm$ 0.15 \\ 
58213.034 & Swift/UVOT    & UVM2 & 18.37 $\pm$ 0.09 \\ 
58215.027 & Swift/UVOT    & UVM2 & 18.50 $\pm$ 0.11 \\ 
58221.218 & Swift/UVOT    & UVM2 & 18.58 $\pm$ 0.16 \\ 
58223.272 & Swift/UVOT    & UVM2 & 18.66 $\pm$ 0.09 \\ 
58228.511 & Swift/UVOT    & UVM2 & 18.91 $\pm$ 0.12 \\ 
58231.965 & Swift/UVOT    & UVM2 & 18.79 $\pm$ 0.12 \\ 
58234.024 & Swift/UVOT    & UVM2 & 18.82 $\pm$ 0.10 \\ 
58237.083 & Swift/UVOT    & UVM2 & 19.02 $\pm$ 0.09 \\ 
58240.142 & Swift/UVOT    & UVM2 & 19.01 $\pm$ 0.09 \\ 
58242.320 & Swift/UVOT    & UVM2 & 19.11 $\pm$ 0.17 \\ 
58249.371 & Swift/UVOT    & UVM2 & 19.14 $\pm$ 0.15 \\ 
58252.154 & Swift/UVOT    & UVM2 & 18.94 $\pm$ 0.09 \\ 
58255.739 & Swift/UVOT    & UVM2 & 18.94 $\pm$ 0.10 \\ 
58261.124 & Swift/UVOT    & UVM2 & 18.76 $\pm$ 0.09 \\ 
58264.045 & Swift/UVOT    & UVM2 & 18.77 $\pm$ 0.08 \\ 
58381.606 & Swift/UVOT    & UVM2 & 19.70 $\pm$ 0.12 \\ 
58396.674 & Swift/UVOT    & UVM2 & 19.79 $\pm$ 0.13 \\ 
58403.571 & Swift/UVOT    & UVM2 & 19.90 $\pm$ 0.26 \\ 
58410.817 & Swift/UVOT    & UVM2 & 19.53 $\pm$ 0.11 \\ 
58417.727 & Swift/UVOT    & UVM2 & 19.79 $\pm$ 0.13 \\ 
58424.426 & Swift/UVOT    & UVM2 & 19.85 $\pm$ 0.18 \\ 
58431.147 & Swift/UVOT    & UVM2 & 19.86 $\pm$ 0.17 \\ 
58438.780 & Swift/UVOT    & UVM2 & 19.73 $\pm$ 0.12 \\ 
58204.530 & Swift/UVOT    & UVW2 & 18.22 $\pm$ 0.07 \\ 
58208.841 & Swift/UVOT    & UVW2 & 18.74 $\pm$ 0.11 \\ 
58210.376 & Swift/UVOT    & UVW2 & 18.49 $\pm$ 0.11 \\ 
58212.638 & Swift/UVOT    & UVW2 & 18.75 $\pm$ 0.16 \\ 
58213.030 & Swift/UVOT    & UVW2 & 18.57 $\pm$ 0.10 \\ 
58215.024 & Swift/UVOT    & UVW2 & 18.55 $\pm$ 0.12 \\ 
58221.215 & Swift/UVOT    & UVW2 & 18.78 $\pm$ 0.09 \\ 
58223.268 & Swift/UVOT    & UVW2 & 18.87 $\pm$ 0.11 \\ 
58228.508 & Swift/UVOT    & UVW2 & 18.90 $\pm$ 0.14 \\ 
58231.962 & Swift/UVOT    & UVW2 & 19.09 $\pm$ 0.15 \\ 
58234.020 & Swift/UVOT    & UVW2 & 19.16 $\pm$ 0.13 \\ 
58237.076 & Swift/UVOT    & UVW2 & 19.13 $\pm$ 0.10 \\ 
58240.135 & Swift/UVOT    & UVW2 & 19.23 $\pm$ 0.10 \\ 
58242.318 & Swift/UVOT    & UVW2 & 19.23 $\pm$ 0.19 \\ 
58249.367 & Swift/UVOT    & UVW2 & 19.10 $\pm$ 0.10 \\ 
58252.148 & Swift/UVOT    & UVW2 & 19.16 $\pm$ 0.11 \\ 
58255.734 & Swift/UVOT    & UVW2 & 19.07 $\pm$ 0.12 \\ 
58261.119 & Swift/UVOT    & UVW2 & 18.79 $\pm$ 0.09 \\ 
58264.038 & Swift/UVOT    & UVW2 & 19.00 $\pm$ 0.09 \\ 
58381.598 & Swift/UVOT    & UVW2 & 19.44 $\pm$ 0.12 \\ 
58396.666 & Swift/UVOT    & UVW2 & 19.59 $\pm$ 0.12 \\ 
58403.568 & Swift/UVOT    & UVW2 & 19.56 $\pm$ 0.20 \\ 
58410.809 & Swift/UVOT    & UVW2 & 19.71 $\pm$ 0.13 \\ 
58417.720 & Swift/UVOT    & UVW2 & 19.45 $\pm$ 0.11 \\ 
58424.422 & Swift/UVOT    & UVW2 & 19.49 $\pm$ 0.16 \\ 
58431.142 & Swift/UVOT    & UVW2 & 19.87 $\pm$ 0.18 \\ 
58438.772 & Swift/UVOT    & UVW2 & 19.56 $\pm$ 0.12 \\ 

\enddata
\tablecomments{Reported magnitudes have the host flux subtracted (see Table~\ref{tab:host_syn}) and are corrected for Galactic extinction. Upper limits are reported at the 5$\sigma$ level. }
\label{tab:photo}
\end{deluxetable}

%% This command is needed to show the entire author+affilation list when
%% the collaboration and author truncation commands are used.  It has to
%% go at the end of the manuscript.
%\allauthors

%% Include this line if you are using the \added, \replaced, \deleted
%% commands to see a summary list of all changes at the end of the article.
%\listofchanges

\end{CJK*}
\end{document}